\documentclass[superscriptaddress, prd, aps,amsmath,amssymb,showpacs,showkeys, onecolumn]{revtex4-2}
\usepackage[dvips]{graphicx,color}
\usepackage{subfig}
\usepackage{enumitem}
\usepackage{times}
\usepackage{float}
\usepackage{xcolor}
\usepackage[%
  colorlinks=true,
  urlcolor=blue,
  linkcolor=red,
  citecolor=blue
]{hyperref}

\begin{document}
\title{Geometric and Thermodynamic Properties of Frolov Black Holes with Topological Defects
}

\author{Ahmad Al-Badawi}
\email[Email: ]{ahmadbadawi@ahu.edu.jo}
\affiliation{Department of Physics, Al-Hussein Bin Talal University 71111, Ma’an, Jordan}

\author{Faizuddin Ahmed}
\email[Email: ]{faizuddinahmed15@gmail.com}

\affiliation{Department of Physics, The Assam Royal Global University, Guwahati, 781035, Assam, India}

\author{ \.{I}zzet Sakall{\i}}
\email[Email: ]{izzet.sakalli@emu.edu.tr}

\affiliation{Physics Department, Eastern Mediterranean University, Famagusta 99628, North Cyprus via Mersin 10, Turkey}

\begin{abstract}
We investigated a modified Frolov black hole (BH) model that incorporates both a global monopole (GM) and a cosmic string (CS) to explore the interplay between non-singular BH regularization and topological defect effects. In our study, we derived a spacetime metric characterized by a regulated core through a length scale parameter $\alpha$ and introduced additional modifications via the GM parameter $\eta$ and the CS parameter $a$, which collectively alter the horizon structure and causal geometry of the BH. We analyzed the thermodynamic properties by deriving expressions for the mass function, Hawking temperature, and entropy, and found that the inclusion of GM and CS significantly deviates the BH entropy from the conventional Bekenstein-Hawking area law, while numerical investigations showed that the shadow radius exhibits contrasting behaviors: the Frolov parameters tend to reduce the shadow size whereas the topological defects enhance it. Furthermore, we examined the dynamics of scalar and electromagnetic perturbations by solving the massless Klein-Gordon equation in the BH background and computed the quasinormal modes (QNMs) using the WKB approximation, which confirmed the BH’s stability and revealed that the oscillation frequencies and damping rates are strongly dependent on the parameters $\alpha$, $q$, $\eta$, and $a$. Our results suggest that the distinct observational signatures arising from this composite BH model may provide a promising avenue for testing modified gravity theories in the strong-field regime.
\end{abstract}


\maketitle

\section{Introduction}\label{sec01}

Over the past century, BHs have transformed from purely mathematical curiosities to observationally confirmed astrophysical objects with the landmark detection of gravitational waves from binary BH mergers by LIGO-Virgo collaborations \cite{isz01,isz02} and the groundbreaking first image of a BH shadow by the Event Horizon Telescope \cite{isz03}. In the framework of classical general relativity (GR), the Kerr-Newman family of BHs, characterized by just three parameters-mass, charge, and angular momentum-stands as the unique solution to Einstein's field equations under the no-hair theorem \cite{isz05,isz06}. However, this theorem faces challenges when considering quantum effects, modified gravity theories, or the presence of topological defects in the BH spacetime.

Among various BH solutions beyond the classical paradigm, the Frolov BH has gained significant attention in recent years \cite{frolov}. Introduced by V. P. Frolov in 2016, this non-singular BH model represents an important advance in addressing the longstanding singularity problem in BH physics. The Frolov BH incorporates a length scale parameter $\alpha$ (related to the Hubble length) that regulates the central singularity through a specific coupling between electromagnetic (EM) and gravitational fields. Unlike conventional Reissner-Nordström (RN) BHs, which contain a physical singularity at $r=0$, the Frolov model employs a radially dependent mass function that effectively smooths out the central region, thereby avoiding the infinite curvature problem \cite{isz07,isz08}. This model is particularly valuable as it preserves the horizon structure and many thermodynamic properties of classical BHs while eliminating the troublesome singularity that conflicts with quantum principles \cite{isz09}.

Topological defects, as relics of symmetry-breaking phase transitions in the early universe, provide another rich arena for extending BH physics beyond standard models. GM, first investigated by Barriola {\it et. al.} \cite{isz10}, represent topological defects arising from the spontaneous breaking of global $O(3)$ symmetry to $U(1)$. The spacetime around a GM exhibits a solid deficit angle and modified gravitational potential, creating distinctive effects on geodesic motion and causal structure \cite{isz11,isz12}. Similarly, cosmic strings are one-dimensional topological defects resulting from the breaking of $U(1)$ symmetry-produce a conical deficit angle in the surrounding spacetime \cite{isz13}. Both GM and CS have been extensively studied for their cosmological implications, from structure formation to gravitational lensing effects \cite{isz14,isz15}, and their incorporation into BH spacetimes introduces remarkable physical behaviors not present in standard solutions.

The integration of these topological defects with BH solutions has led to compelling composite spacetime structures. Various authors extended BHs with GM in the literature \cite{isz16,isz17,isz18}. Similarly, BHs threaded by CS have been investigated for their modified properties compared to their standard counterparts \cite{isz19,isz20}. Such composite solutions are not merely theoretical constructs but represent plausible astrophysical configurations that could form through various evolutionary pathways in the early universe \cite{isz21}. The combined Frolov BH with dual topological defects (GM and CS), as studied in this paper, therefore represents a physically motivated extension that incorporates both non-singular behavior and topological effects into a unified spacetime structure.

BH thermodynamics constitutes one of the most profound connections between gravity, quantum theory, and information science. Since Hawking's seminal discovery that BH emit thermal radiation with a temperature proportional to their surface gravity \cite{isz22,isz23}, the thermodynamic properties of BHs have been extensively analyzed. The four laws of BH thermodynamics, formulated by Bardeen, Carter, and Hawking \cite{isz24}, establish direct analogies between BH physics and conventional thermodynamics, with the BH mass, horizon area, and surface gravity corresponding to energy, entropy, and temperature, respectively. These thermodynamic properties serve as crucial probes of a BH's stability, phase transitions, and information-theoretic characteristics \cite{isz25,isz26}. For modified BH solutions like the Frolov BH with GM and CS, the thermodynamic analysis becomes particularly revealing, as it can identify unique behaviors that distinguish these solutions from standard BHs and potentially lead to observable signatures \cite{isz27}.

The BH shadow-a dark region against a bright background as viewed by a distant observer-represents one of the few directly observable features of BHs. The shadow's size and shape depend sensitively on the BH's mass, spin, and the underlying spacetime geometry \cite{isz28}. Following the groundbreaking observations by the Event Horizon Telescope collaboration \cite{isz03}, BH shadow studies have evolved from theoretical curiosities to powerful tools for testing GR and alternative gravity theories \cite{isz29,isz30}. The shadow of non-standard BH, such as the Frolov BH with GM and CS, can exhibit distinctive features that potentially allow for observational discrimination between different theoretical models \cite{isz31,isz32}. The shadow calculation involves determining the photon sphere-a region where light rays can orbit the BH in unstable circular orbits-which marks the boundary between photons that escape to infinity and those captured by the BH \cite{isz33}.

The study of perturbations around BH spacetime provides crucial insights into their stability and response to external disturbances. Scalar, EM, and gravitational perturbations induce characteristic oscillations known as QNMs, which have specific frequencies and damping times determined by the BH parameters \cite{isz34,isz35}. These QNMs not only probe the stability of the underlying spacetime but also carry distinctive "fingerprints" of the BH's fundamental properties \cite{isz36}. In the context of gravitational wave astronomy, QNMs manifest in the ring-down phase following BH mergers, offering a direct observational window into BH physics \cite{isz37,isz38}.

The motivation for studying the Frolov BH with GM and CS stems from several compelling factors. First, the synthesis of non-singular BH models with topological defects represents a natural theoretical progression that addresses both the singularity problem and the influence of early universe phase transitions on BH formation and evolution. Second, the increasing precision of astrophysical observations, particularly in gravitational wave astronomy and BH imaging, creates opportunities for testing extended BH models against observational data. Third, the composite solutions-from modified thermodynamic behavior to distinctive shadow characteristics and perturbation spectra-serve as important test beds for exploring potential signatures of quantum gravity or modified gravity theories in the strong-field regime \cite{isz39,isz40}. In this paper, we conduct a comprehensive investigation of the Frolov BH with GM and CS, focusing on its geometric, thermodynamic, and perturbative properties. We systematically analyze how the four key parameters of the model-the length scale parameter $\alpha$, the charge $q$, the GM parameter $\eta$, and the CS parameter $a$-influence various physical characteristics of the spacetime. Our analysis reveals several notable features: (1) the presence of GM and CS significantly modifies the horizon structure and thermodynamic stability, (2) the shadow radius exhibits opposing trends with respect to the Frolov parameters versus the topological defect parameters, and (3) the QNM spectrum shows distinctive patterns that potentially allow for observational discrimination of this model from ordinary GR BHs. Throughout our analysis, we emphasize the physical implications of these findings and their potential observational signatures.

The paper is organized as follows: In Section \ref{sec02}, we introduce the Frolov BH with GM and CS, examining its metric structure and horizon properties. We study the dynamics of photons in the geometry of Frolov BH with GM and CS in Section \ref{sec03}. Section \ref{sec04} is devoted to a detailed analysis of the BH's thermodynamic characteristics, including mass, temperature, entropy, and heat capacity functions. In Section \ref{sec05}, we investigate the BH shadow, calculating the photon sphere and shadow radius for various parameter configurations. Section \ref{sec06} focuses on scalar and EM perturbations, deriving the effective potential and computing the QNM frequencies using the WKB approximation. Finally, in Section \ref{sec07}, we summarize our findings and discuss their implications for theoretical and observational BH physics. Throughout the paper, we adopt natural units where $G = c = \hbar = k_B = 1$.

\section{FEATURES OF Frolov BH with GM and CS}\label{sec02}

The Frolov BH represents a significant advancement in addressing the central singularity problem that plagues classical BH solutions \cite{isz41,isz42}. When combined with topological defects such as GM and CS, the resulting spacetime exhibits rich physical properties that merit detailed investigation. In this section, we analyze the fundamental features of the Frolov BH with GM and CS, focusing on its metric structure, action, and horizon characteristics.

The action describing Frolov BH with GM and CS incorporates both EM and gravitational components, along with terms that represent the energy contributions of the topological defects. This action is expressed as follows \cite{isz46}:

\begin{equation}
S_{\mathrm{GM}}=\int d^4 x \sqrt{-g}\left[\frac{1}{16 \pi}(R-2 \Lambda)-\frac{1}{4} F_{\mu \nu} F^{\mu \nu}-\frac{1}{2} \partial_\mu \phi^a \partial^\mu \phi^a-\frac{\lambda}{4}\left(\phi^a \phi^a-\eta^2\right)^2\right].
\end{equation}

When the scalar fields settle into a "hedgehog" configuration at large $r$, namely $\phi^a \propto x^a$, the vacuum expectation value $\eta$ induces a solid-angle deficit in the metric. This deficit appears in the metric function $\mathcal{F}(r)$ as a modification by the term $1-8 \pi \eta^2$. In addition, the Maxwell term $\frac{1}{4} F_{\mu \nu} F^{\mu \nu}$ supplies the usual Coulomb-type charge $q$ to the solution, while the inclusion of the cosmological constant $\Lambda$ introduces a large distance scale $\alpha$, which is related by $\alpha \sim 1/\sqrt{\Lambda}$ \cite{isz10,isz11}. In many treatments, the condition $\phi^a \phi^a=\eta^2$ is enforced in the broken-symmetry vacuum by taking the limit $\lambda \rightarrow \infty$, thereby leaving behind an effective "monopole" hair. As a result, one obtains a static and spherically symmetric Frolov BH with GM and CS, described by the following line element \cite{frolov}:
\begin{equation}
    ds^2=-\mathcal{F}(r)\,dt^2+\frac{dr^2}{\mathcal{F}(r)}+r^2\,(d\theta^2+\sin^2\theta\,d\phi^2),\label{bb1}
\end{equation}
where the metric function $\mathcal{F}(r)$ (in natural units) is 
\begin{equation}
    \mathcal{F}(r)=1-a-8\,\pi\,\eta^2-\frac{(2\,Mr-q^2)r^2}{r^4+(2\,Mr+q^2)\alpha^2},    \label{bb2}
\end{equation}
where $a$ represents the CS parameter \cite{isz47} and $\alpha$ is known as the Hubble length (length scale parameter) \cite{isz50,isz51}, serving as a free parameter of the model. The Hubble length displays itself as Universal hair and is restricted by $\alpha\leq \sqrt{16/27}M$.  

The metric given in Eq. (\ref{bb1}) captures several important physical elements: (1) the parameter $a$ represents the CS contribution, manifesting as a deficit angle in the spacetime \cite{isz10}; (2) the term $8\pi\eta^2$ reflects the GM's influence, where $\eta$ is related to the energy scale of spontaneous symmetry breaking that formed the monopole \cite{isz10}; (3) the fraction term encapsulates the Frolov BH's regular behavior, with $q$ representing the electric charge $(0\leq q \leq 1)$ and $M$ the mass parameter.

Several well-known solutions in GR can be recovered from this metric by taking appropriate limits. For instance: 
\begin{enumerate}[label=(\roman*)]
\item When $\alpha \to 0$, $a \to 0$, and $\eta \to 0$, one recovers the RN metric, which has been extensively studied in the literature.
\item When $\alpha \to 0$, $q \to 0$, and $\eta \to 0$, we recover the Letelier BH solution \cite{isz47}.
\item When $a \to 0$, $q \to 0$, and $\alpha \to 0$, one can find the GM spacetime \cite{isz10}.
\item When $a \to 0$ and $\eta \to 0$, we get back the original Frolov BH solution \cite{frolov}.
\end{enumerate}

The asymptotic behavior of the metric function at the origin and at extreme distances are given by:
\begin{eqnarray}
    &&\mathcal{F}(r)=1-a-8\,\pi\,\eta^2-\frac{2\,M}{r}+\frac{q^2}{r^2}+\alpha^2\,\mathcal{O}(r^{-4})\quad (r \to \infty),\label{bb2a}\\
    &&\mathcal{F}(r)=1-a-8\,\pi\,\eta^2+\frac{r^2}{\alpha^2}+\mathcal{O}(r^6)\quad (r \to 0).\label{bb2b}
\end{eqnarray}

The relation given in Eq. (\ref{bb2b}) confirms that the corresponding metric remains regular at the origin $r=0$, where its curvature is of the order of $\alpha^{-2}$. However, at large distances $r \to \infty$, the metric function $\mathcal{F}(r)$ in Eq. (\ref{bb2a}) begins to deviate from the RN metric due to the influence of dual topological defects-namely, the CS and GM contributions characterized by the parameters $(a, \eta)$. This deviation indicates the presence of a conical geometry, implying that while the spacetime is locally flat, it is not globally flat. As a result, the spacetime is asymptotically non-flat.

It is also important to emphasize that the presence of both the GM parameter $\eta$ and the CS parameter $a$ fundamentally alters the spacetime geometry compared to standard BH solutions.  In the regime $r\gg\alpha,M,q$, the metric function no longer approaches unity but instead tends to the constant value $1 - a - 8\pi\eta^2$, manifesting a conical deficit angle and a non‐Minkowskian, asymptotically non‐flat structure.  Although the Kretschmann scalar $R^{\mu\nu\lambda\sigma}\,R_{\mu\nu\lambda\sigma}$ vanishes at spatial infinity $r \to \infty$, a detailed expansion shows it diverges as $r^{-4}$ at the origin $r=0$, confirming that a true curvature singularity persists despite the regular Frolov‐core form enhanced by the dual topological defects.  This interplay between a modified far‐field geometry and an unavoidable central divergence has profound consequences for gravitational lensing, light deflection, and orbital motion in the far‐field region.

\begin{figure}[ht!]
    \centering
    \includegraphics[width=0.5\linewidth]{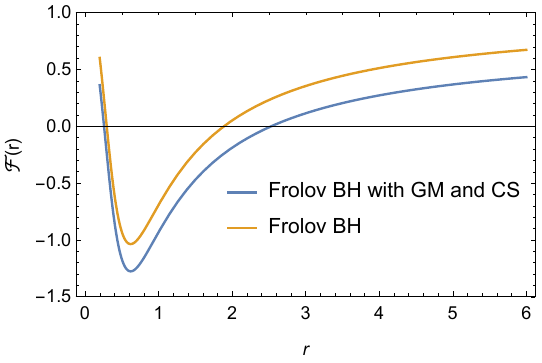}
    \caption{Lapse function $\mathcal{F}(r)$ as a function of $r$. It shows that the BH has two horizons. Here, we use $a=0.2$, $\alpha=0.2$, $\eta=0.2$, and $q=0.6$.}
    \label{lapse1}
\end{figure}

To better understand the metric function (\ref{bb2}), we display it as a function of $r$, as illustrated in Fig. \ref{lapse1}. The graph demonstrates that there are only two horizons for the Frolov BH with GM and CS: the inner horizon ($r_{h-}$) and the outer horizon ($r_{h+}$).

The horizons can be determined using the condition:  
\begin{equation}
    1-a-8\,\pi\,\eta^2-\frac{(2\,Mr-q^2)r^2}{r^4+(2\,Mr+q^2)\alpha^2}=0. \label{hor1} 
 \end{equation}
 
Unlike the standard RN BH, where horizons can be determined analytically, Eq. (\ref{hor1}) has no closed-form solution due to the complex interplay of the Frolov parameter $\alpha$ with the topological defect parameters $a$ and $\eta$. The presence of the length scale parameter $\alpha$ fundamentally alters the causal structure of the spacetime, ensuring its regularity at the origin while preserving the horizon structure characteristic of charged BHs.

\begin{figure}[h]
   \includegraphics[scale=0.65]{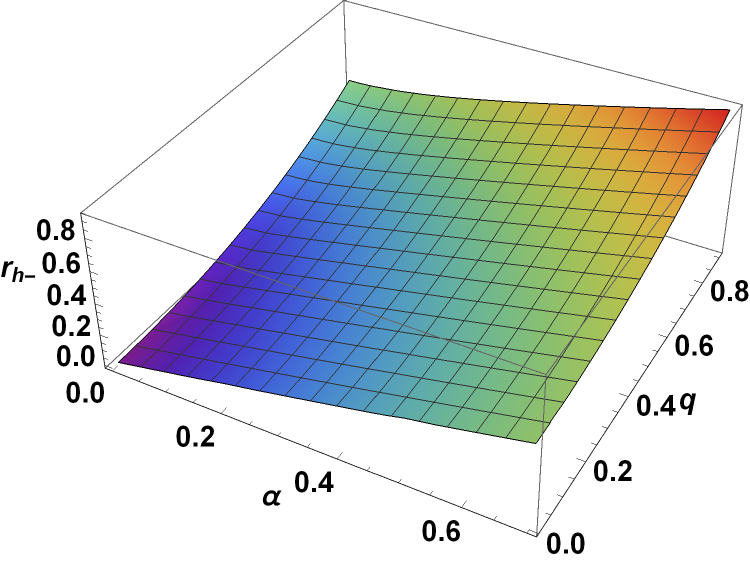}\quad
   \includegraphics[scale=0.65]{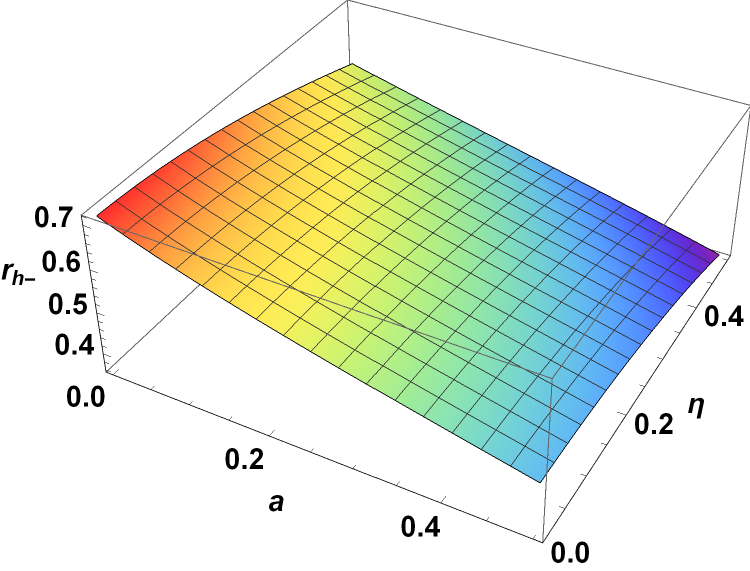}
    \caption{Variation of the inner horizon for different values of $\alpha$ and $q$ (left), and for different values of $\eta$ and $a$ (right).}
    \label{Horiz1}
\end{figure}

\begin{figure}[h]
   \includegraphics[scale=0.65]{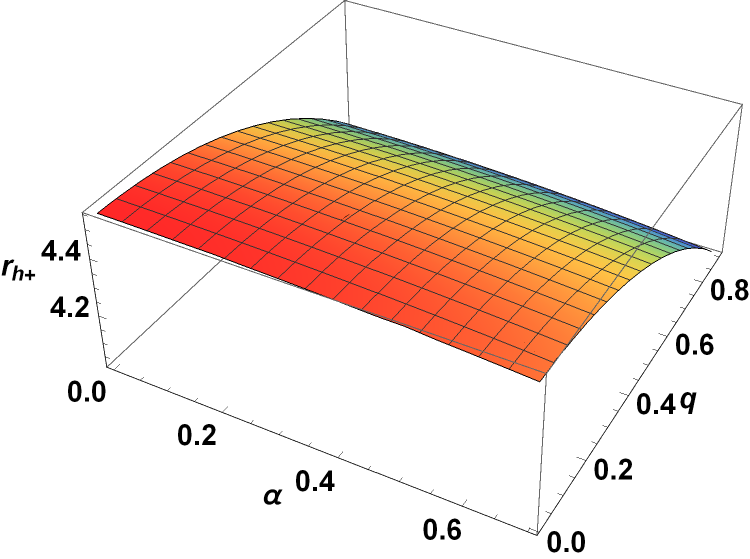}\quad
   \includegraphics[scale=0.65]{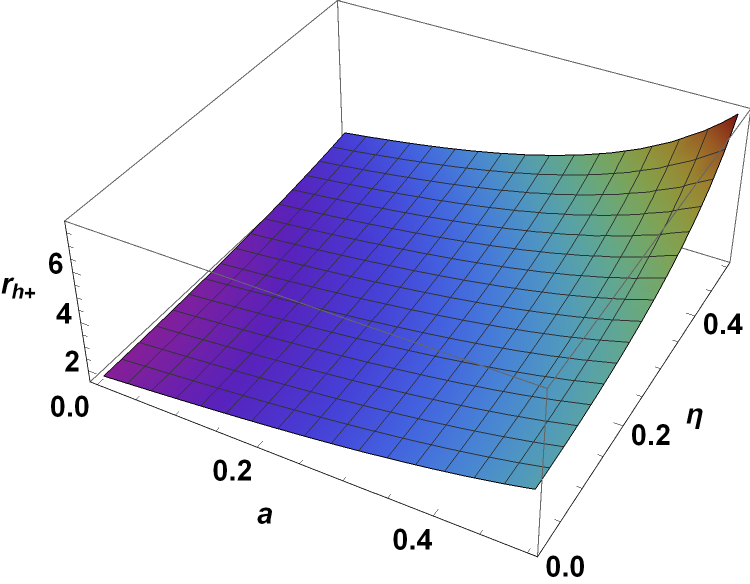}
    \caption{Variation of the outer horizon for different values of $\alpha$ and $q$ (left), and for different values of $\eta$ and $a$ (right).}
    \label{Horiz2}
\end{figure}

We employ numerical methods to calculate the event and Cauchy horizons. Tables \ref{tab3a} and \ref{tab4a} present the numerical results for these two horizons under various parameter combinations. Figure \ref{Horiz1} depicts the inner horizon and shows how the parameters $\alpha$, $q$, $a$, and $\eta$ influence it, while Figure \ref{Horiz2} shows the corresponding effects on the outer horizon.

\begin{center}
\begin{tabular}{|c |c| c|c|c|c|c|c|c|c|}
 \hline   & \multicolumn{3}{|c|}{ $q=0.2$} & \multicolumn{3}{|c|}{ $0.4$} &  \multicolumn{3}{|c|}{ $0.6$} \\ \hline 
 $\alpha $ & $r_{h-}$ & $r_{h+}$ & $r_{h+}-r_{h-}$ & $r_{h-}$ & $r_{h+}$ & $r_{h+}-r_{h-}$ & $r_{h-}$ & $r_{h+}$ & $r_{h+}-r_{h-}$ \\ \hline
$0.2$ & $0.154185$ & $4.52142$ & $4.36724$ & $0.20719$ & $4.45982$ & $4.25263$ & $0.295159$ & $4.3531$ & $4.05794$ \\ 
$0.4$ & $0.294424$ & $4.50949$ & $4.21507$ & $0.352427$ & $4.44722$ & $4.09479$ & $0.445525$ & $4.3392$ & $3.89368$ \\ 
$0.6$ & $0.439386$ & $4.48932$ & $4.04993$ & $0.501542$ & $4.42587$ & $3.92433$ & $0.60167$ & $4.3156$ & $3.71393$\\ 
 \hline
\end{tabular}
\captionof{table}{Numerical results for the inner and outer horizons of Frolov BH with GM and CS for various values of parameters $\alpha$ and $q$. Here $a=0.4$ and $\eta=0.4$.} \label{tab3a}
\end{center}

\begin{center}
\begin{tabular}{|c |c| c|c|c|c|c|c|c|c|}
 \hline   & \multicolumn{3}{|c|}{ $\eta=0.2$} & \multicolumn{3}{|c|}{ $0.4$} &  \multicolumn{3}{|c|}{ $0.6$}  \\ \hline
$a$ & $r_{h-}$ & $r_{h+}$ & $r_{h+}-r_{h-}$ & $r_{h-}$ & $r_{h+}$ & $r_{h+}-r_{h-}$ & $r_{h-}$ & $r_{h+}$ & $r_{h+}-r_{h-}$ \\ \hline
$0.2$ & $0.582236$ & $2.37048$ & $1.78824$ & $0.52899$ & $2.88963$ & $2.36064$ & $0.445525$ & $4.3392$ & $3.89368$ \\ 
$0.4$ & $0.495324$ & $3.34944$ & $2.85412$ & $0.445525$ & $4.3392$ & $3.89368$ & $0.357059$ & $8.14422$ & $7.78716$ \\ 
$0.6$ & $0.411624$ & $5.35737$ & $4.94575$ & $0.357059$ & $8.14422$ & $7.78716$ & $0.230508$ & $49.8192$ & $49.5887$\\ 
 \hline
\end{tabular}
\captionof{table}{Numerical results for the inner and outer horizons of Frolov BH with GM and CS for various values of parameters $a$ and $\eta$. Here $\alpha=0.4$ and $q=0.6$.} \label{tab4a}
\end{center}

Analysis of these figures reveals several important patterns. Figure \ref{Horiz1} indicates that $\alpha$ and $q$ have similar effects, with the inner horizon increasing with both (see left panel). However, the inner horizon shrinks substantially when $a$ and $\eta$ rise (see right panel). Conversely, Figure \ref{Horiz2} shows that increasing $\alpha$ and $q$ values decrease the outer horizon radius (left panel), while the outer horizon expands significantly as $a$ and $\eta$ increase (right panel). This opposing behavior between inner and outer horizons under parameter variation illustrates the complex interplay between regularization and topological effects in determining the causal structure of the spacetime.

\section{Dynamics of Photons in FROLOV BH WITH GM AND CS} \label{sec03}

In this section, we study motions of photon light around the Frolov BH solution with GM and CS and analyze how various parameters, such as the GM and CS parameters, and electric charge alters the dynamics of photon light. Geodesic motions of photon light gives us the useful information about the photon sphere radius and BH shadow, stability of circular orbits under the influence of the gravitational field produced by BHs. Geodesics motions in various singular and regular BHs have widely been studied in the literature (see, for example, Refs. \cite{AHEP1,AHEP2,NPB,CJPHY,AHEP3,AHEP4,EPJC} and related references therein). The effective potential for null geodesics around a BH is a tool to describe the behavior of light in curved spacetime near a BH. It helps us to understand the motion of light around BHs, particularly in regions where the spacetime is highly curved, such as near the event horizon. The effective potential provides insight into whether photons will be able to escape from the BH's gravitational pull (if they are outside the photon sphere) or whether they will be captured by the BH. With the help of the effective potential, one can determine the photon sphere radius including the BH shadow and other.  

The motion of photon light is described by $ds^2=0$. For the given spacetime (\ref{bb1}), we obtain
\begin{equation}
    -\mathcal{F}(r)\,\left(\frac{dt}{d\tau}\right)^2+\frac{1}{\mathcal{F}(r)}\,\left(\frac{dr}{d\tau}\right)^2+r^2\,\left(\frac{d\theta}{d\tau}\right)^2+r^2\,\sin^2 \theta\,\left(\frac{d\phi}{d\tau}\right)^2=0,\label{mm1}
\end{equation}
where $\tau$ represents an affine parameter.

Considering the geodesics motion in the equatorial plane defined by $\theta=\pi/2$ and $\frac{d\theta}{d\tau}=0$, we find the following conserved quantities 
\begin{equation}
    \mathrm{E}=\mathcal{F}(r)\,\left(\frac{dt}{d\tau}\right),\quad\quad \mathrm{L}=r^2\,\frac{d\phi}{d\tau},\label{mm2}
\end{equation}
where $\mathrm{E}$ is the conserved energy associated with temporal coordinate and $\mathrm{L}$ is the conserved angular momentum.

With these, we can rewrite Eq. (\ref{mm1}) after rearranging as
\begin{equation}
    \left(\frac{dr}{d\tau}\right)^2+V_\text{eff}(r)=\mathrm{E}^2\label{mm3}
\end{equation}
which is equivalent to the one-dimensional equation of motion with $V_\text{eff}(r)$ is the effective potential given by
\begin{equation}
    V_\text{eff}(r)=\frac{\mathrm{L}^2}{r^2}\,\mathcal{F}(r)=\frac{\mathrm{L}^2}{r^2}\,\left(1-a-8\,\pi\,\eta^2-\frac{(2\,M\,r-q^2)\,r^2}{r^4+(2\,M\,r+q^2)\,\alpha^2}\right).\label{mm4}
\end{equation}

From expression given in Eq. (\ref{mm4}), it is clear that the effective potential for null geodesics is influenced by several factors. These include the energy-scale of the symmetry-breaking $\eta$, the CS parameter $a$, the length scale parameter $\alpha$, and the electric charge $q$. Additionally, the conserved angular momentum $\mathrm{L}$ and BH mass $M$ alters the effective potential.

\begin{figure}[ht!]
    \centering
    \subfloat[$a=0.1,\,\eta^2=0.1$]{\centering{}\includegraphics[width=0.4\linewidth]{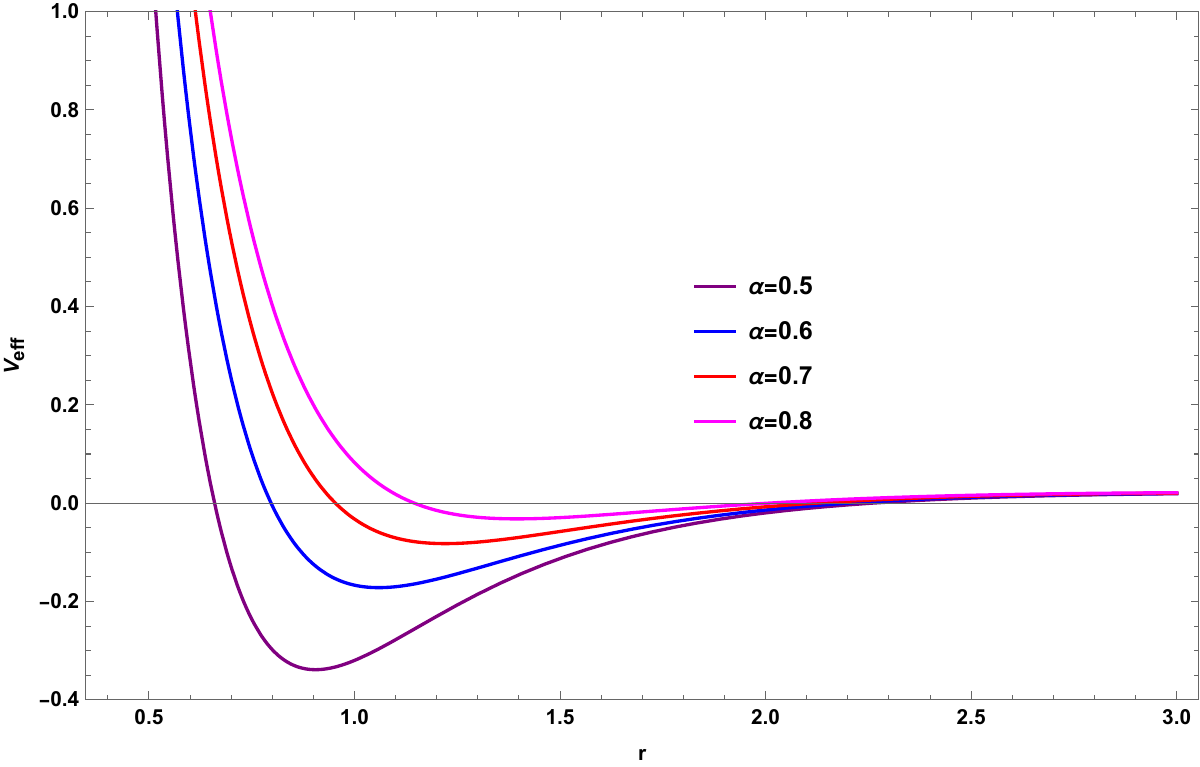}}\quad\quad
    \subfloat[$\alpha=0.5,\,\eta^2=0.1$]{\centering{}\includegraphics[width=0.4\linewidth]{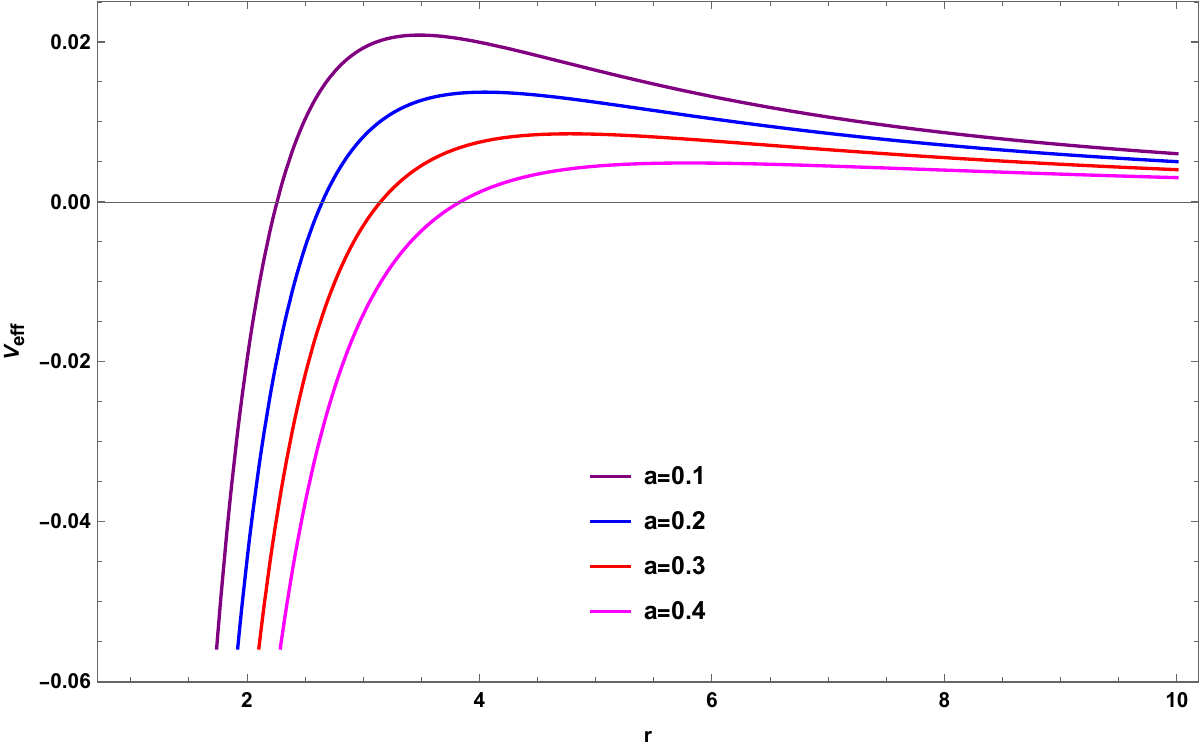}}\\
    \subfloat[$a=0.1,\,\alpha=0.5$]{\centering{}\includegraphics[width=0.4\linewidth]{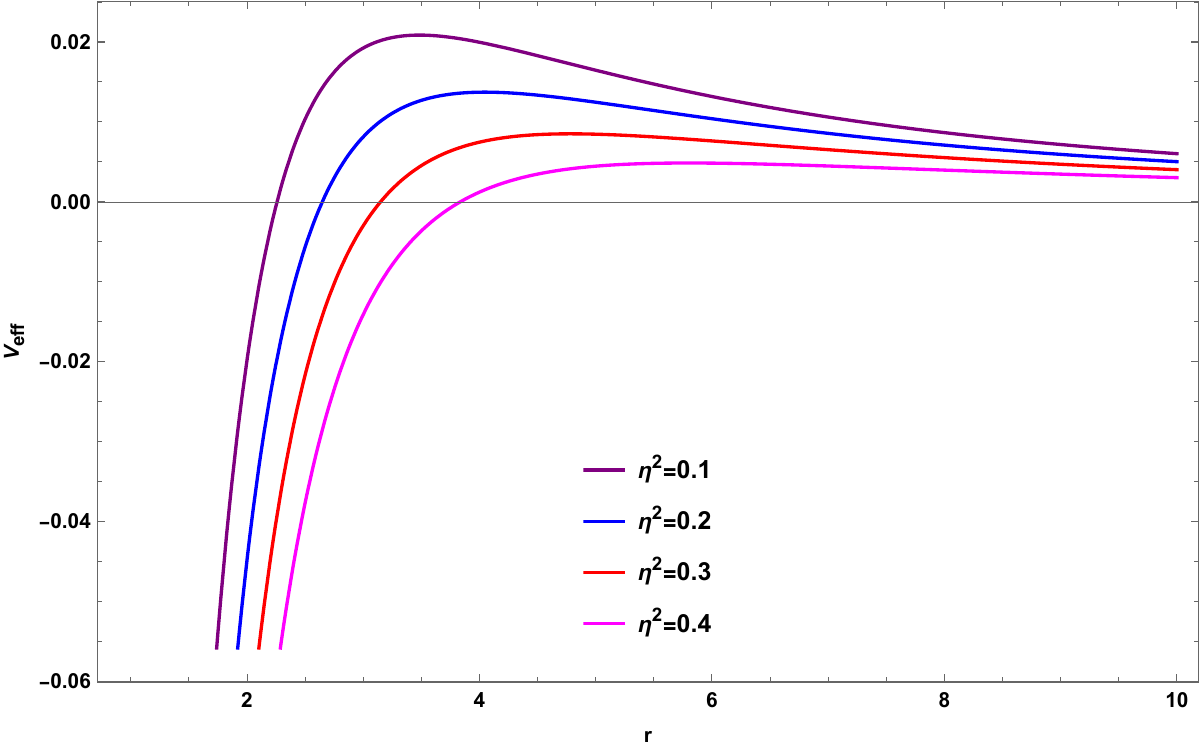}}\quad\quad
    \subfloat[$\alpha=0.5$]{\centering{}\includegraphics[width=0.4\linewidth]{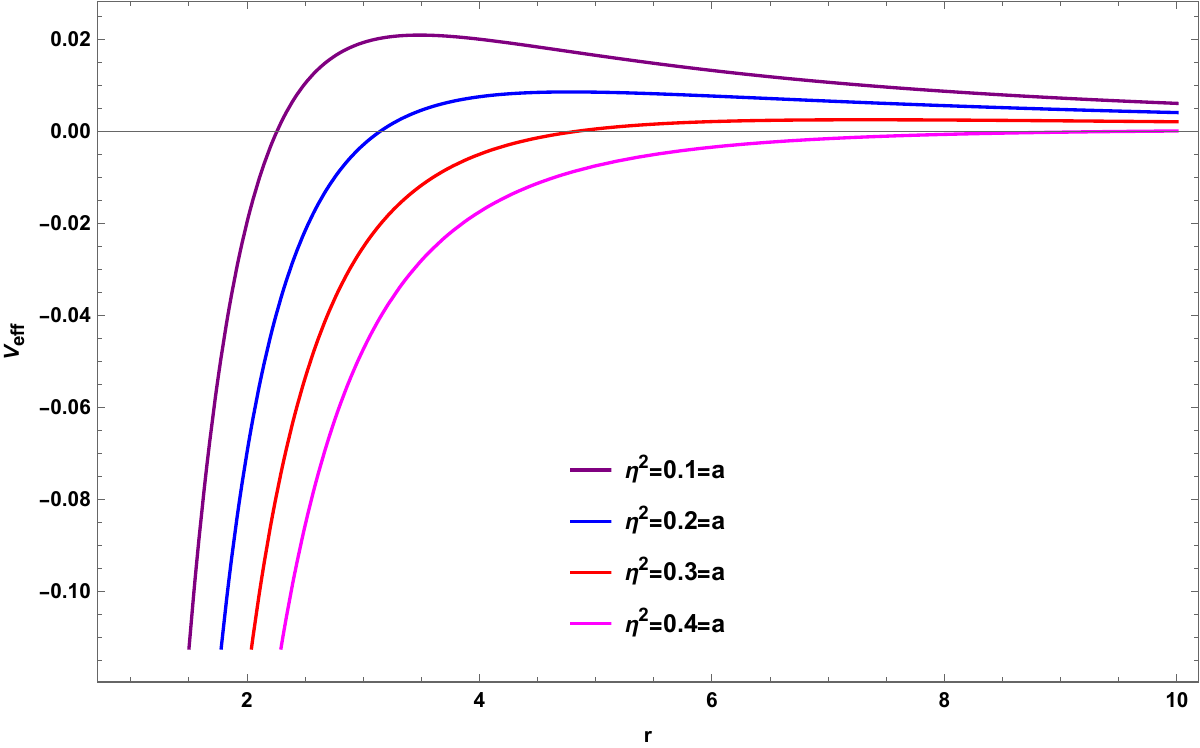}}\\
    \subfloat[$a=0.1$]{\centering{}\includegraphics[width=0.4\linewidth]{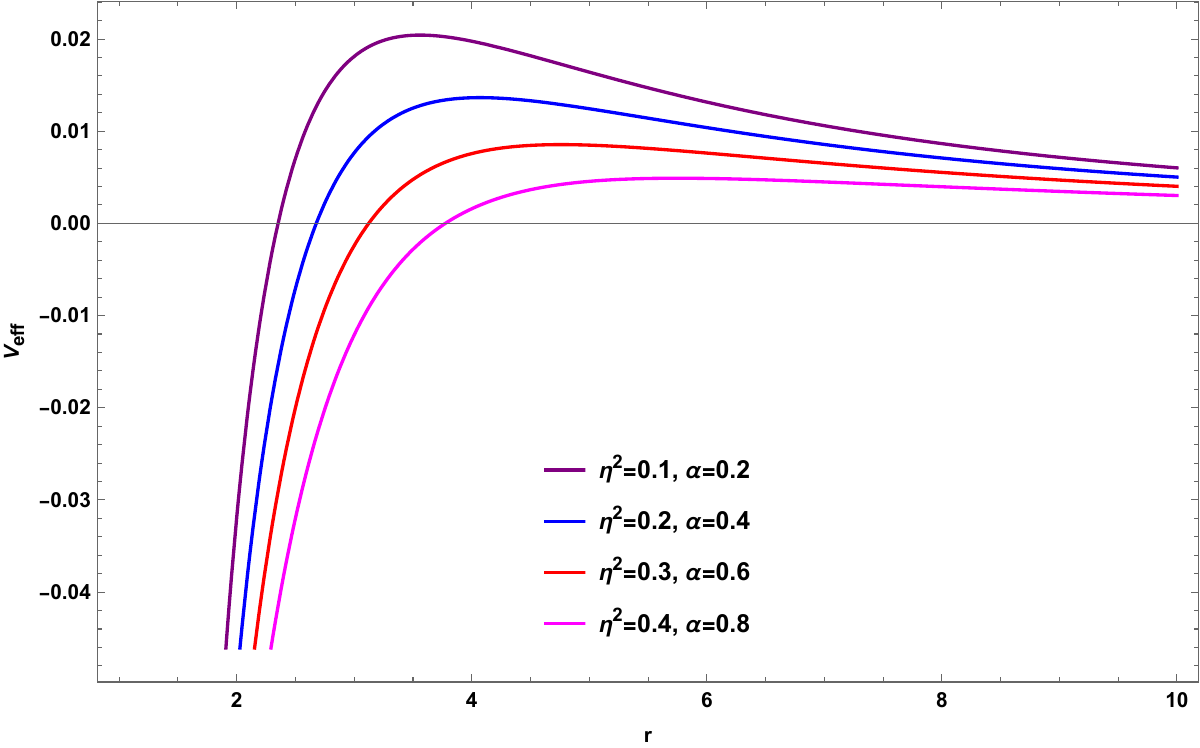}}\quad\quad
    \subfloat[]{\centering{}\includegraphics[width=0.4\linewidth]{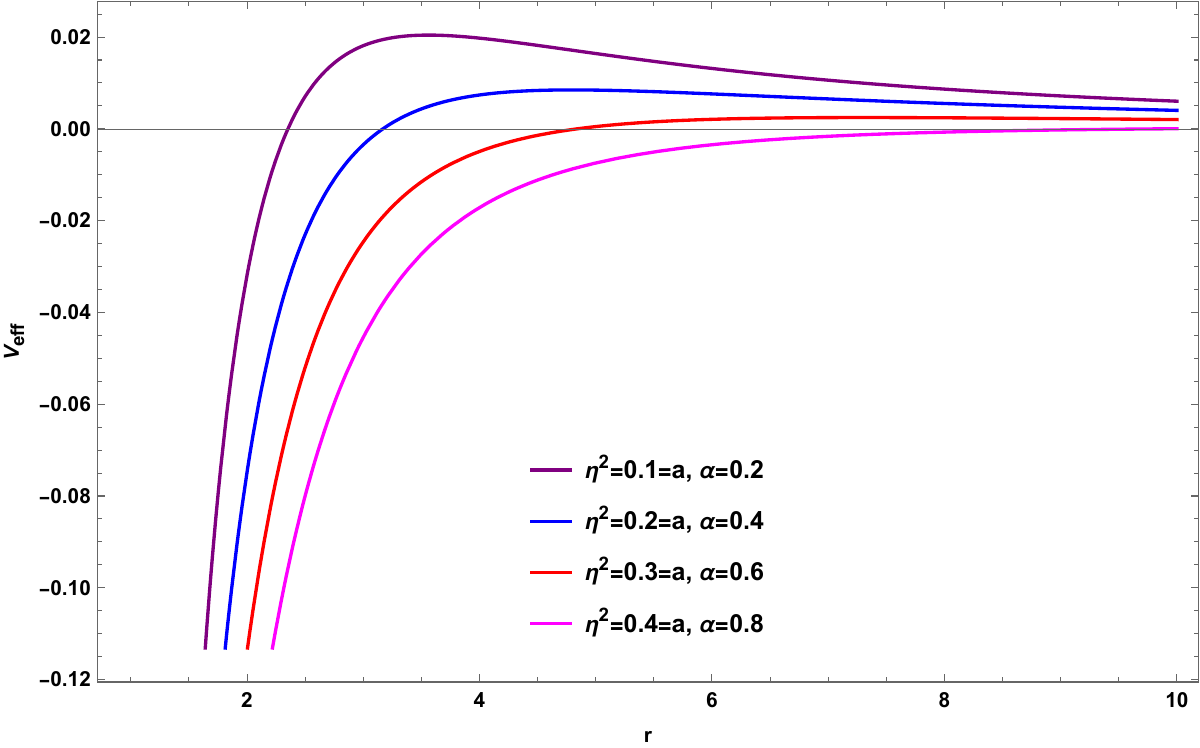}}
    \caption{The behavior of the effective potential for null geodesics varying different parameter $a, \eta$ and $\alpha$. Here, we set $M=1$, $\mathrm{L}=1$ and $q=0.5$.}
    \label{fig:potential}
\end{figure}

In Figure \ref{fig:potential}, we generate the effective potential for null geodesics by varying the values of the CS parameter $a$, the symmetry-breaking energy-scale parameter $\eta$, and the length scale parameter $\alpha$, all individually and in combination. In panel (a) of Figure \ref{fig:potential}, we observe that increasing the length scale parameter from $\alpha=0.5$ leads to a corresponding rise in the effective potential with increasing radial coordinate $r$. In contrast, panels (b) through (f) show that increasing the CS parameter $a$, the energy scale parameter $\eta$, or their combinations with the length scale parameter $\alpha$, results in a decrease in the effective potential as $r$ increases. Throughout this Figure, we fix the BH mass $M=1$, the angular momentum $\mathrm{L}=1$, and the electric charge to $q=0.5$. These parameters collectively influence the shape and behavior of the effective potential for null geodesics in the vicinity of the considered BH.

For circular motions around the BH vicinity, we have the conditions $\frac{dr}{d\tau}=0$ and $\frac{d^2r}{d\tau^2}=0$, we yields
the following relations
\begin{equation}
    \mathrm{E}^2=V_\text{eff}(r)=\frac{\mathrm{L}^2}{r^2}\,\left(1-a-8\,\pi\,\eta^2-\frac{(2\,M\,r-q^2)\,r^2}{r^4+(2\,M\,r+q^2)\,\alpha^2}\right).\label{mm5}
\end{equation}
The above relation gives us the critical impact parameter for photon light given by
\begin{equation}
    \frac{1}{\beta^2_c}=\frac{1-a-8\,\pi\,\eta^2}{r^2}-\frac{(2\,M\,r-q^2)}{r^4+(2\,M\,r+q^2)\,\alpha^2}.\label{mm6}
\end{equation}

From expression given in Eq. (\ref{mm6}), it is clear that the critical impact parameter for photon particles is influenced by several factors. These include the energy-scale of the symmetry-breaking $\eta$, the CS parameter $a$, the length scale parameter $\alpha$, and the electric charge $q$. Additionally, the BH mass $M$ alters this impact parameter.

And the second relation is $V'_\text{eff}(r)=0$ which gives us the photon sphere radius $r=r_\text{ph}$ satisfies the following relation:
\begin{equation}
    r\,\mathcal{F}'(r)-2\,\mathcal{F}(r)=0\Rightarrow 1-a-8\,\pi\,\eta^2+\frac{M\,r^3}{r^4+(2\,M\,r+q^2)\,\alpha^2}+\frac{(M\,\alpha^2\,q^2-2\,\alpha^2\,M^2\,r+2\,q^2\,r^3-4\,M\,r^4)\,r^3}{[r^4+(2\,M\,r+q^2)\,\alpha^2]^2}.\label{mm7}
\end{equation}

Now, we determine the force on photons under the gravitational field produced by the BH and show how various parameters affect the motion of light in the vicinity of the BH. Using the effective potential for null geodesics given in Eq. (\ref{mm4}), we can determine the force on photons as
\begin{equation}
    \mathrm{F}_\text{ph}=-\frac{1}{2}\,\frac{dV_\text{eff}}{dr}=\frac{\mathrm{L}^2}{r^3}\,\left[1-a-8\,\pi\,\eta^2+\frac{M\,r^3}{r^4+(2\,M\,r+q^2)\,\alpha^2}+\frac{(M\,\alpha^2\,q^2-2\,\alpha^2\,M^2\,r+2\,q^2\,r^3-4\,M\,r^4)\,r^3}{[r^4+(2\,M\,r+q^2)\,\alpha^2]^2} \right].\label{mm8}
\end{equation}

\begin{figure}[ht!]
    \centering
    \subfloat[$a=0.1,\,\eta^2=0.1$]{\centering{}\includegraphics[width=0.4\linewidth]{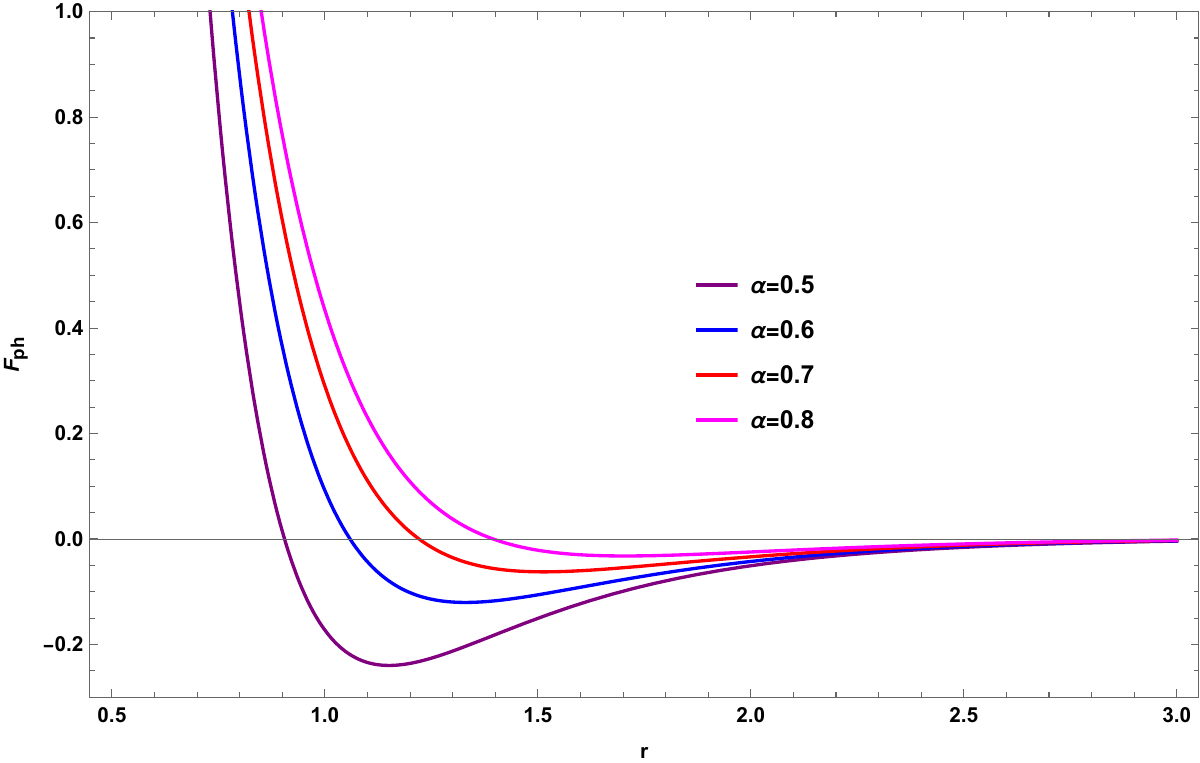}}\quad\quad
    \subfloat[$\alpha=0.5,\,\eta^2=0.1$]{\centering{}\includegraphics[width=0.4\linewidth]{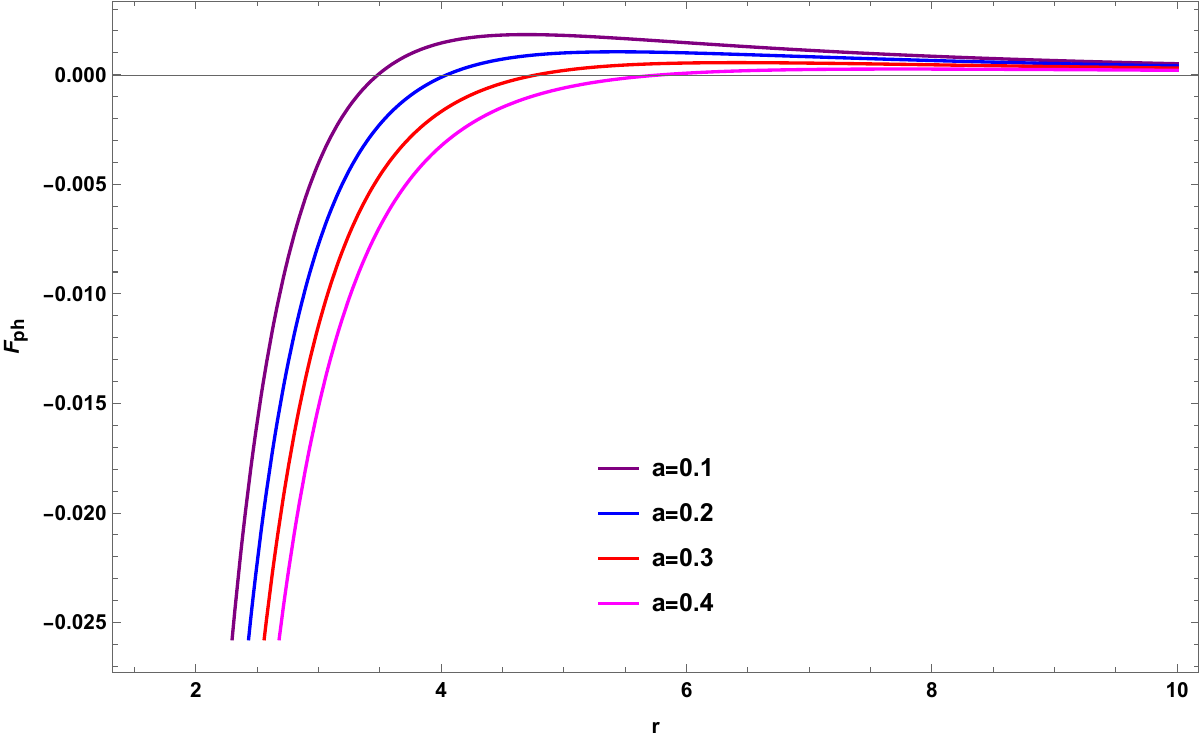}}\\
    \subfloat[$a=0.1,\,\alpha=0.5$]{\centering{}\includegraphics[width=0.4\linewidth]{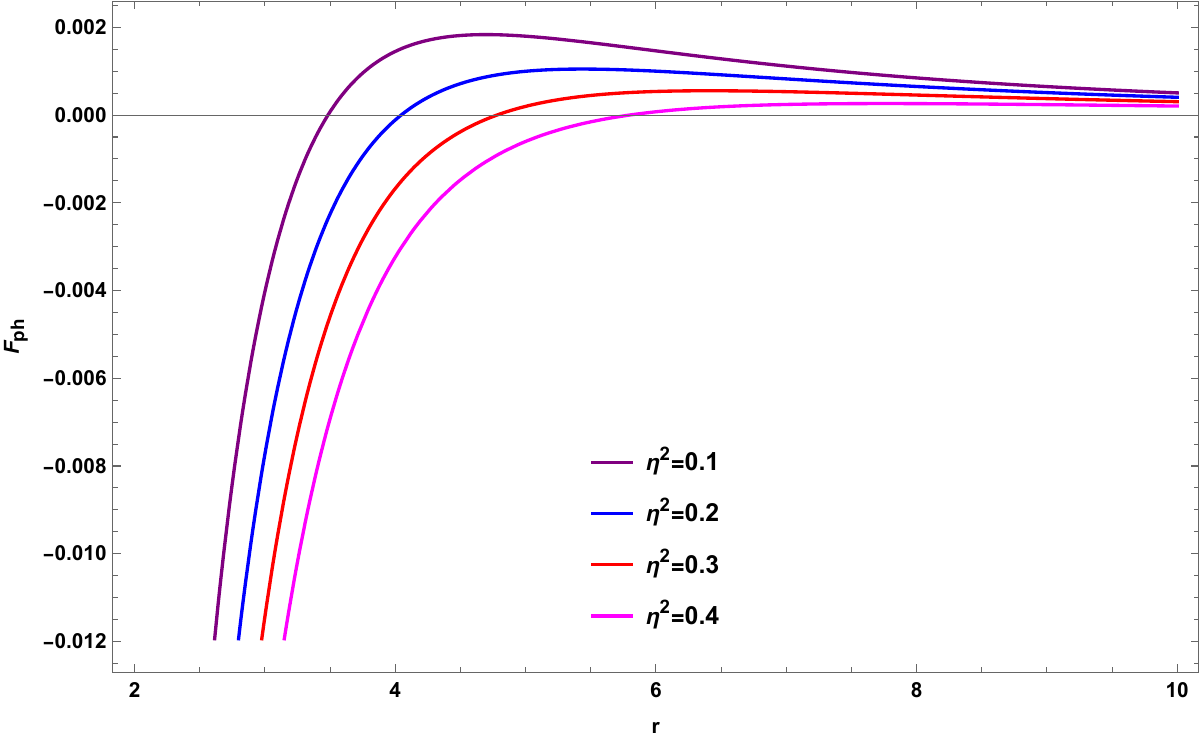}}\quad\quad
    \subfloat[$\alpha=0.5$]{\centering{}\includegraphics[width=0.4\linewidth]{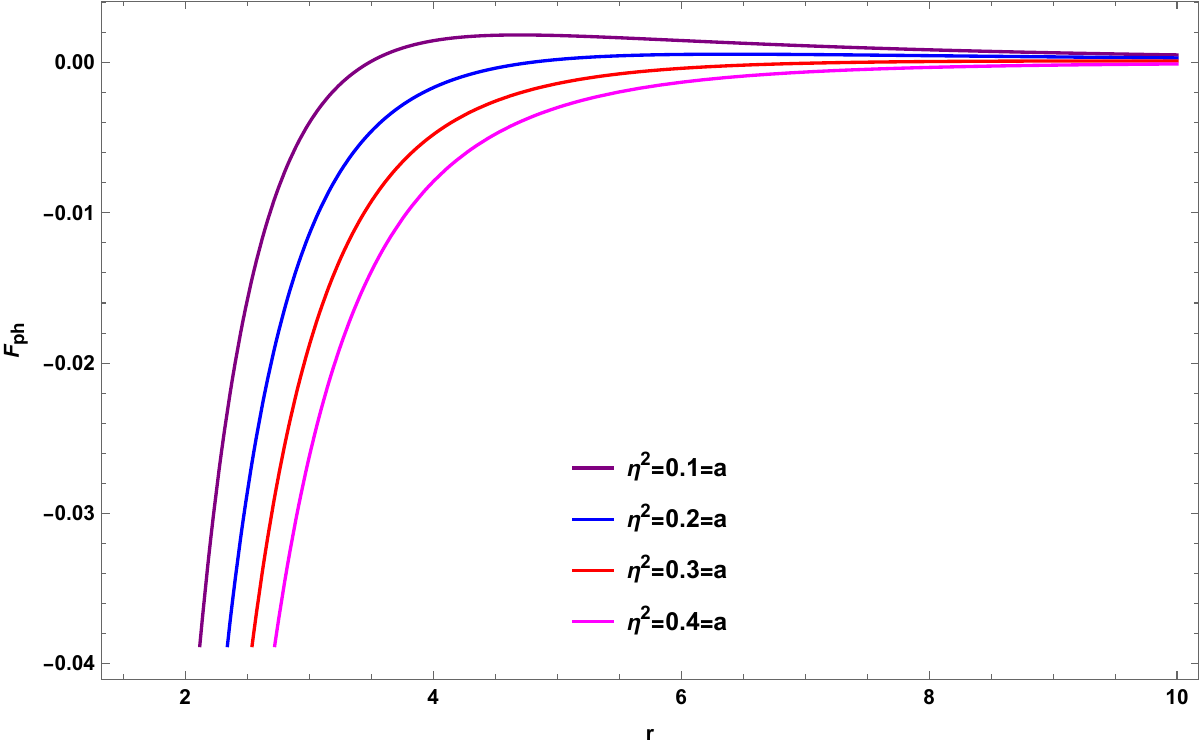}}\\
    \subfloat[$a=0.1$]{\centering{}\includegraphics[width=0.4\linewidth]{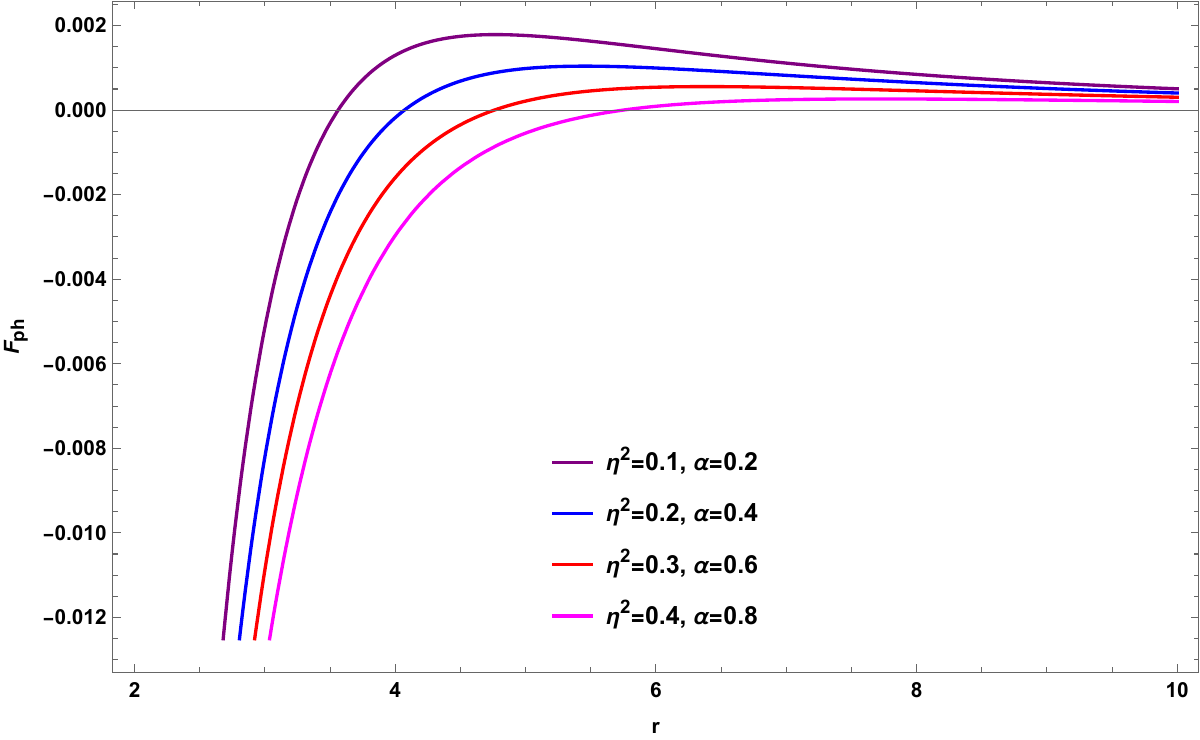}}\quad\quad
    \subfloat[]{\centering{}\includegraphics[width=0.4\linewidth]{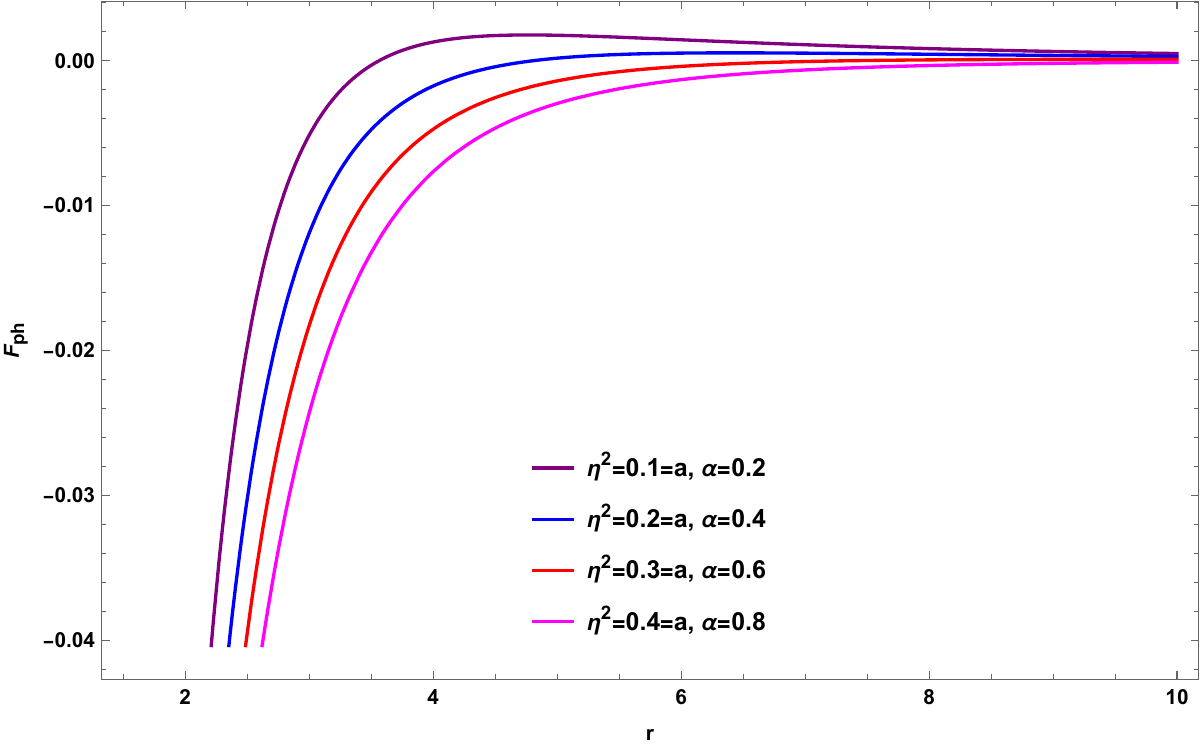}}
    \caption{The behavior of force on photon particles by varying different parameter $a, \eta$ and $\alpha$. Here, we set $M=1$, $\mathrm{L}=1$ and $q=0.5$.}
    \label{fig:force}
\end{figure}

From expression given in Eq. (\ref{mm8}), it is clear that the force on photon particles depend on various factors. These include the energy-scale of the symmetry-breaking $\eta$, the CS parameter $a$, the length scale parameter $\alpha$, and the electric charge $q$. Additionally, the conserved angular momentum $\mathrm{L}$, and BH mass $M$ also alters the force acting on photon particles under the gravitational field produced by the selected charged BH. 

In Figure \ref{fig:force}, we illustrate the behavior of the force acting on photon particles by varying the CS parameter $a$, the symmetry-breaking energy scale parameter $\eta$, and the length scale parameter $\alpha$, both individually and in combination. In panel (a), we observe that increasing the length scale parameter from $\alpha = 0.5$ leads to a corresponding increase in the force with the radial coordinate $r$. This suggests that larger values of $\alpha$ enhance the gravitational field, resulting in a stronger attraction of photon particles. In contrast, panels (b) through (f) show that increasing the CS parameter $a$, the energy scale parameter $\eta$, or their combinations with $\alpha$, leads to a decrease in the force as $r$ increases. This indicates that larger values of these parameters weaken the effective gravitational field generated by the charged BH, thereby reducing the attraction experienced by photon particles. Throughout the analysis, we fix the BH mass to $M = 1$, the angular momentum to $\mathrm{L} = 1$, and the electric charge to $q = 0.5$. Together, these parameters govern the dynamics of photon particles, influencing whether they are captured by or escape from the gravitational field of the BH. 

To differentiate force with the known result, we consider the limit where $\alpha=0$, that is absence of the length scalar parameter in the selected BH solution, zero CS parameter $a=0$, and zero the energy-scale parameter $\eta=0$. In that limit, the chosen BH solution reduces to RN-metric, where the metric function reduces to as $\mathcal{F}\to f=1-\frac{2\,M}{r}+\frac{q^2}{r^2}$. 

Therefore, the expression of force on photon particles in this limit from Eq. (\ref{mm8}) becomes 
\begin{equation}
    \mathrm{F}_\text{ph}=\frac{\mathrm{L}^2}{r^3}\,\left(1-\frac{3\,M}{r}+\frac{2\,q^2}{r^2} \right).\label{mm9}
\end{equation}

Thereby, comparing the expression of force given in Eq. (\ref{mm8}) with that of Eq. (\ref{mm9}), it is clear that the current result gets modification (either increased or decreased) by the energy-scale of the symmetry-breaking $\eta$, the CS parameter $a$, and length scale parameter $\alpha$.   

\begin{figure}[ht!]
    \centering
    \subfloat[$a=0.1,\,\eta^2=0.1$]{\centering{}\includegraphics[width=0.4\linewidth]{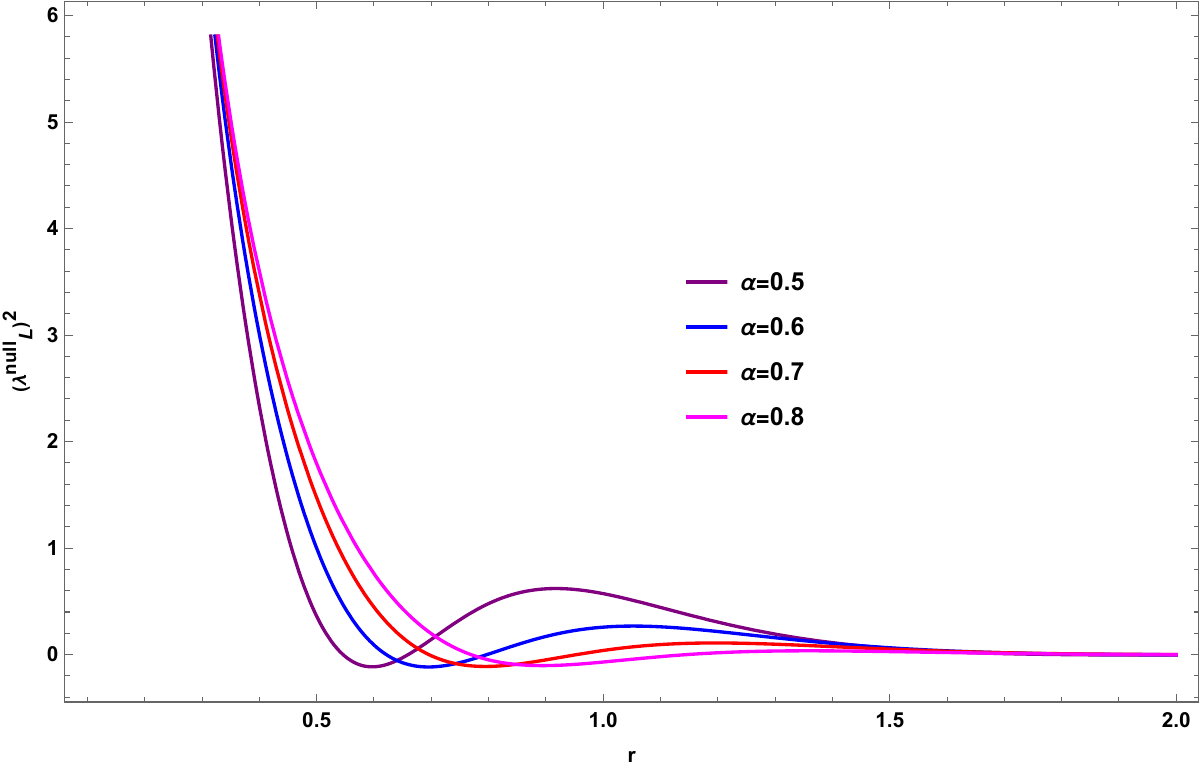}}\quad\quad
    \subfloat[$\alpha=0.5,\,\eta^2=0.1$]{\centering{}\includegraphics[width=0.42\linewidth]{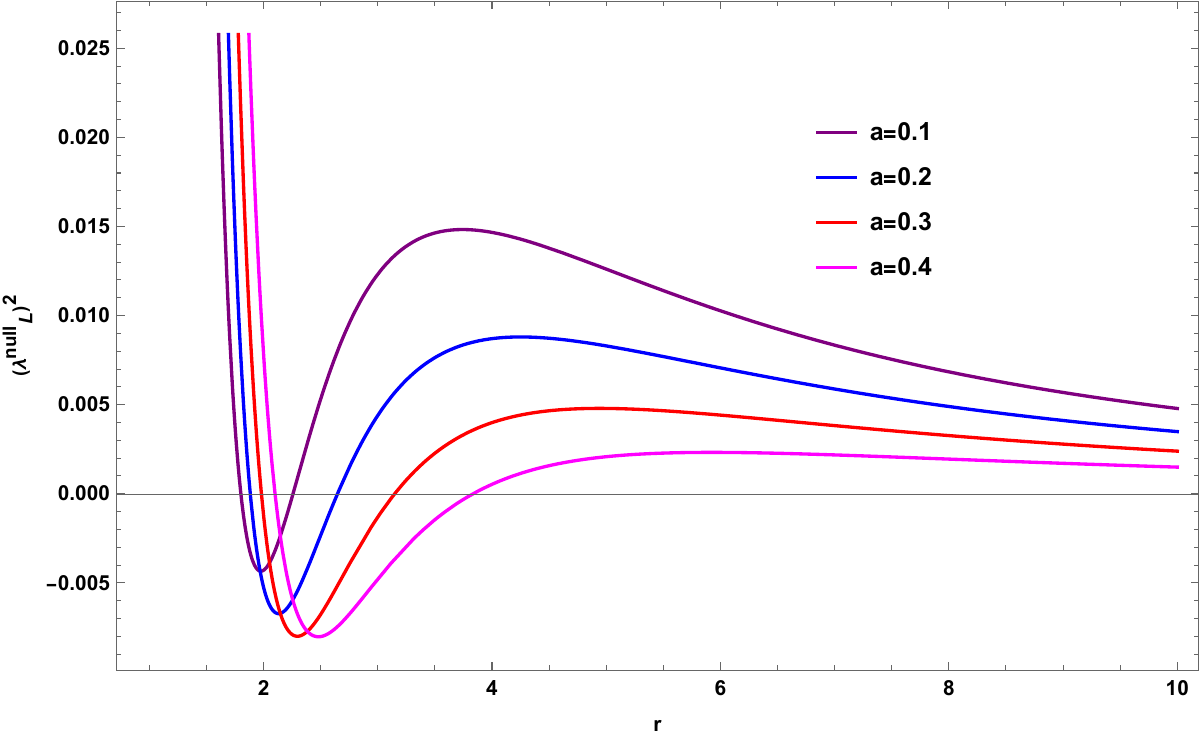}}\\
    \subfloat[$a=0.1,\,\alpha=0.5$]{\centering{}\includegraphics[width=0.4\linewidth]{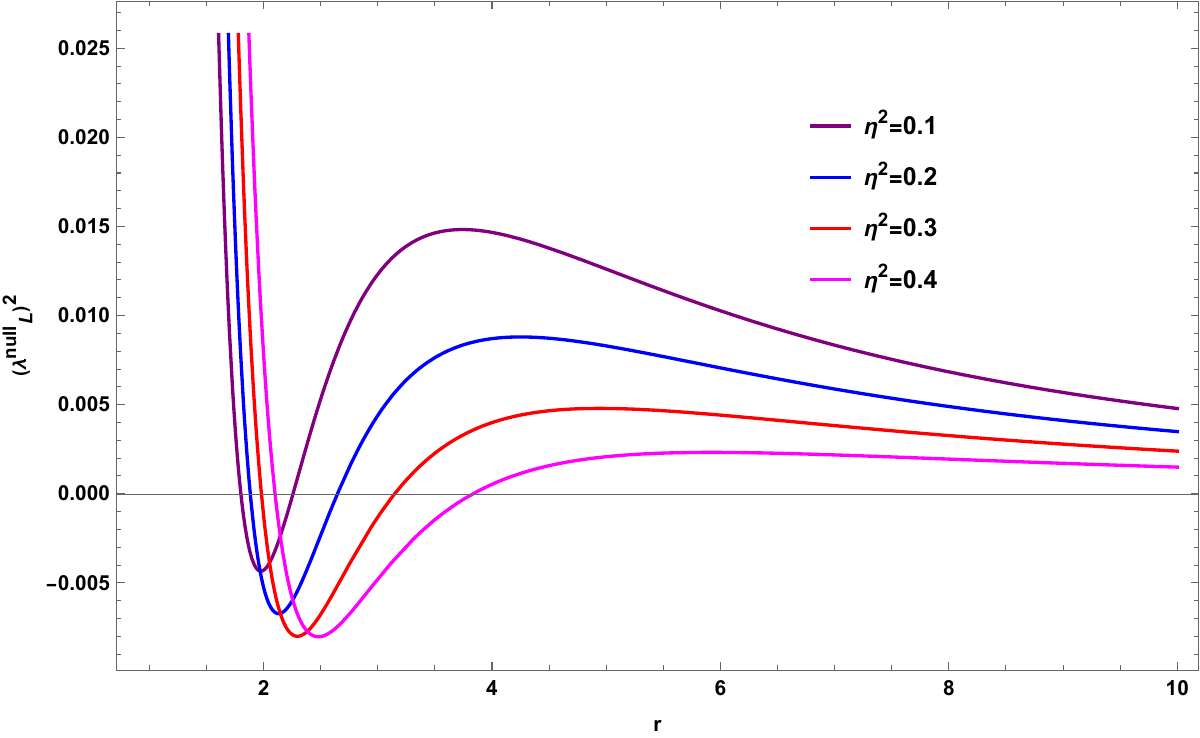}}\quad\quad
    \subfloat[$\alpha=0.5$]{\centering{}\includegraphics[width=0.4\linewidth]{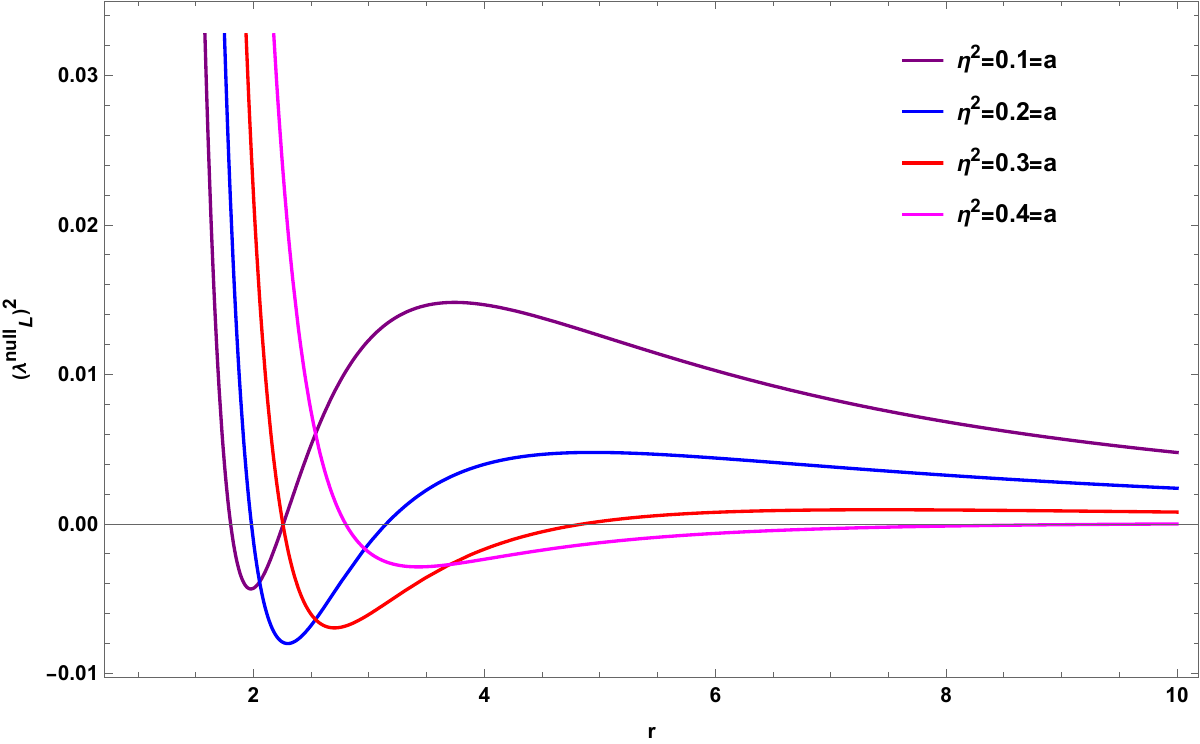}}\\
    \subfloat[$a=0.1$]{\centering{}\includegraphics[width=0.4\linewidth]{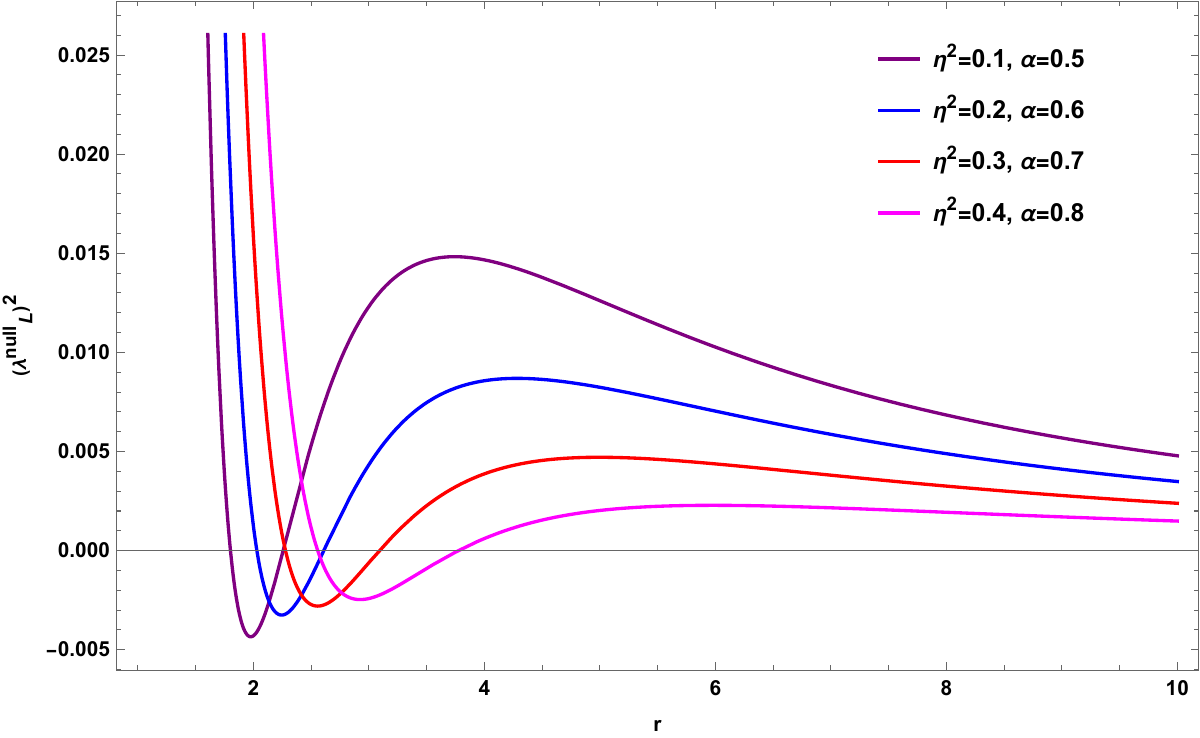}}\quad\quad
    \subfloat[]{\centering{}\includegraphics[width=0.4\linewidth]{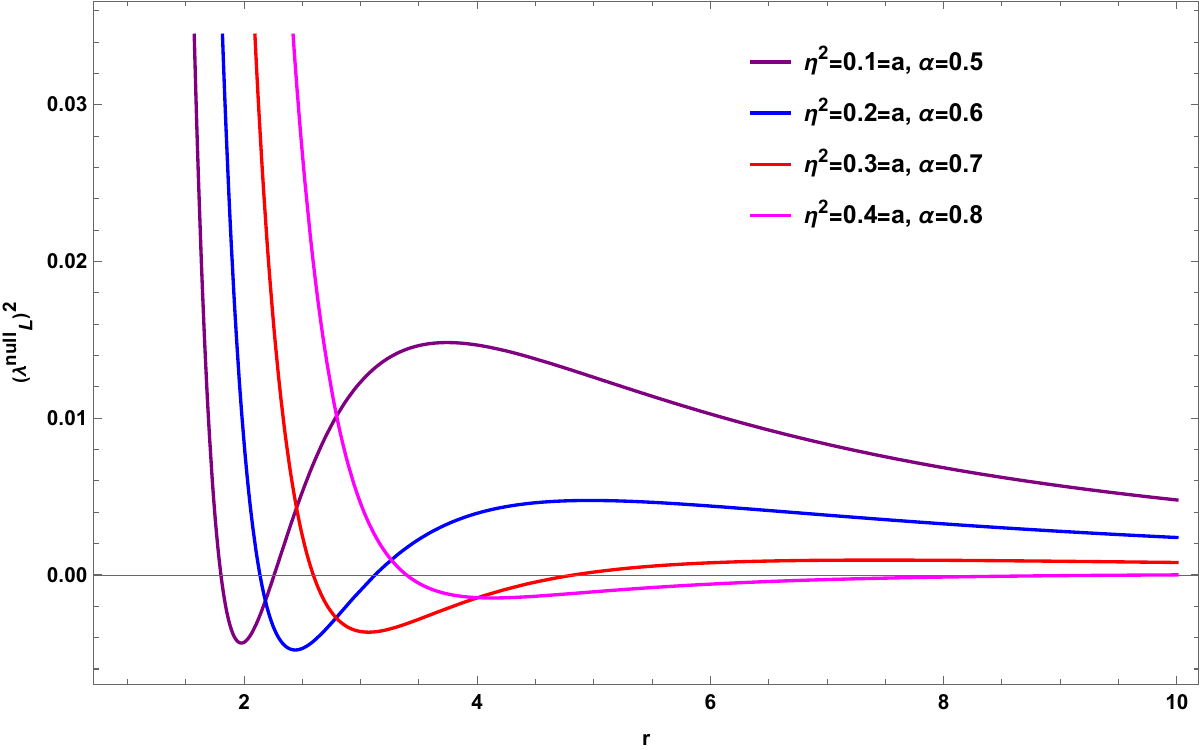}}
    \caption{The behavior of square of the Lyapunov exponent by varying different parameter $a, \eta$ and $\alpha$. Here, we set $M=1$, $\mathrm{L}=1$ and $q=0.5$.}
    \label{fig:lyapunov}
\end{figure}

We now turn our attention to an important physical quantity that determines the stability of circular null orbits-the Lyapunov exponent. This exponent characterizes the (in)stability of the orbits and is derived from the second derivative of the effective potential for null geodesics, as follows: \cite{VC}
\begin{equation}
    \lambda^\text{null}_L=\sqrt{-\frac{V''_\text{eff}(r)}{2\,\dot{t}^2}},\label{mm10}
\end{equation}
where $\dot{t}$ is given in Eq. (\ref{mm2}).

Using the effective potential for null geodesics given in Eq. (\ref{mm4}), we find the Lyapunov exponent for circular null geodesics in terms of the metric function $\mathcal{F}(r)$ as follows:
\begin{equation}
    \lambda^\text{null}_L=\sqrt{\mathcal{F}(r)\,\left(\frac{\mathcal{F}}{r^2}-\frac{\mathcal{F''}}{2}\right)},\label{mm11}
\end{equation}
where we have used the relation $r\,\mathcal{F}'(r)=2\,\mathcal{F}(r)$.

Substituting the metric function given in Eq. (\ref{bb2}), we find the Lyapunov exponent for circular null orbits as (setting $8\,\pi=1$)
\begin{eqnarray}
    \lambda^\text{null}_L&=&\Bigg[-\frac{\left(1-a-8\,\pi\,\eta^2-\frac{(2\,M\,r-q^2)\,r^2}{r^4+(2\,M\,r+q^2)\,\alpha^2}\right)}{r^2\, \left(r^4 + (q^2 + 2\, M\, r)\, \alpha^2\right)^3}\,\Big\{q^6\,\alpha^6\, (-1 + a + \eta^2) +q^4\, r\, \alpha^2\, \Big(-14\, r^5 - 8\, M\, r^2\, \alpha^2 +3\, r^3\, \alpha^2\, (-1 + a + \eta^2)\nonumber\\
    &&+6\, M\, \alpha^4\, (-1 + a + \eta^2)\Big)+ q^2\, r^2\, \Big(2\, r^8 + 12\, M\, r^5\, \alpha^2 - 8\, M^2\, r^2\, \alpha^4+3\, r^6\, \alpha^2\, (-1 + a + \eta^2) \nonumber\\
    &&+ 
    12\, M\, r^3\, \alpha^4\, (-1 + a + \eta^2) + 
    12\, M^2\, \alpha^6\, (-1 + a + \eta^2)\Big)+r^3\, \Big(36\, M^2\, r^5\, \alpha^2 + r^9\, (-1 + a + \eta^2)\nonumber\\
    && + 
    6\, M\, r^6\, \alpha^2\, (-1 + a + \eta^2)  + 
    12\, M^2\, r^3\, \alpha^4\, (-1 + a + \eta^2) + 
    8\, M^3\, \alpha^6\, (-1 + a + \eta^2)\Big)\Big\}\Bigg]^{1/2}.
\end{eqnarray}

In Figure \ref{fig:lyapunov}, we present the behavior of the square of the Lyapunov exponent associated with circular null geodesics, by varying the CS parameter $a$, the symmetry-breaking energy scale parameter $\eta$, and the length scale parameter $\alpha$, both individually and in combination. In panel (a), we observe that increasing the length scale parameter from $\alpha = 0.5$ results in positive values of the Lyapunov exponent squared as a function of the radial coordinate $r$. This indicates that larger values of $\alpha$ enhance the gravitational field, leading to exponential divergence or convergence of nearby circular null trajectories-signifying chaotic behavior and unstable circular orbits. In contrast, panels (b) through (f) show that increasing the CS parameter $a$, the energy scale $\eta$, or their combinations with $\alpha$, causes fluctuations in the square of the Lyapunov exponent with respect to $r$. This behavior suggests that larger values of these parameters reduce the effective gravitational strength of the charged BH, leading to regions where circular orbits may alternate between stability and instability, depending on the radial distance $r=r_c$, where $r_c$ is the radius of the circular null orbits. Throughout the analysis, we fix the BH mass to $M = 1$, the angular momentum to $\mathrm{L} = 1$, and the electric charge to $q = 0.5$. 

Finally, the geodesics angular velocity (called coordinate velocity) is defined by \cite{VC}
\begin{equation}
    \omega=\frac{\dot{\phi}}{\dot{t}}=\frac{\sqrt{\mathcal{F}(r)}}{r}.\label{mm12}
\end{equation}
Here, we have used the relation $\mathrm{E}=\frac{\mathrm{L}}{r}\,\sqrt{\mathcal{F}}$.

Substituting the metric function given in Eq. (\ref{bb2}), we find the geodesics angular velocity
\begin{equation}
    \omega=\sqrt{\frac{1-a-8\,\pi\,\eta^2}{r^2}-\frac{(2\,M\,r-q^2)}{r^4+(2\,M\,r+q^2)\,\alpha^2}}.\label{mm13}
\end{equation}

From expression given in Eq. (\ref{mm13}), it is clear that null geodesics angular velocity in the equatorial plane depend on various factors. These include the energy-scale of the symmetry-breaking $\eta$, the CS parameter $a$, and the length scale parameter $\alpha$. Additionally, the BH mass $M$, and charge $q$ also alters the angular velocity under the given gravitational field of the charged BH.

\section{Thermodynamics OF FROLOV BH WITH GM AND CS} \label{sec04}

The thermodynamic analysis of BHs has become a cornerstone in understanding the deep connections between gravity, quantum theory, and information science \cite{isz23,isz24}. For modified BH solutions like the Frolov BH with GM and CS, examining thermodynamic properties reveals distinctive behaviors that differentiate these solutions from classical BHs and potentially lead to observable signatures \cite{isz63}. In this section, we conduct a comprehensive thermodynamic analysis of the Frolov BH with GM and CS, investigating how the interplay between the length scale parameter $\alpha$, charge $q$, GM parameter $\eta$, and CS parameter $a$ influences various thermodynamic quantities.

Let us begin by deriving the mass function for the Frolov BH with GM and CS. By applying the horizon condition $\mathcal{F}(r_+)=0$ at the event horizon radius $r_{+}$, we obtain:
\begin{equation}
   M_H=\frac{q^2(r^2_+-\alpha^2(a+\eta^2-1))-r^4_+(a+\eta^2-1)}{2r^2_+(r^2_+-\alpha^2+a\alpha^2+\alpha^2\eta^2)}. \label{mass1}  
\end{equation}

The mass function in Eq. (\ref{mass1}) differs significantly from that of standard BH solutions due to the complex interplay between the Frolov parameter $\alpha$ and the topological defect parameters $a$ and $\eta$ \cite{isz64}. Unlike the ADM mass in asymptotically flat spacetimes, the mass $M_H$ here incorporates the contribution of the background geometry modified by the GM and CS. In specific limits, we recover known results: without GM and CS ($a=0=\eta$), Eq. (\ref{mass1}) reduces to the mass of the original Frolov BH \cite{isz64}, and when $\alpha=0$, it reduces to the mass of the RN BH as $M_H=\frac{q^{2}+r_{+}^{2}}{2\,r_{+}}$.

\begin{figure}[ht!]
\begin{center}
\includegraphics[scale=0.9]{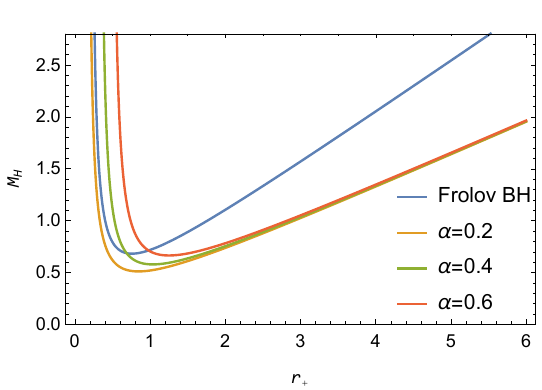}\quad
\includegraphics[scale=0.9]{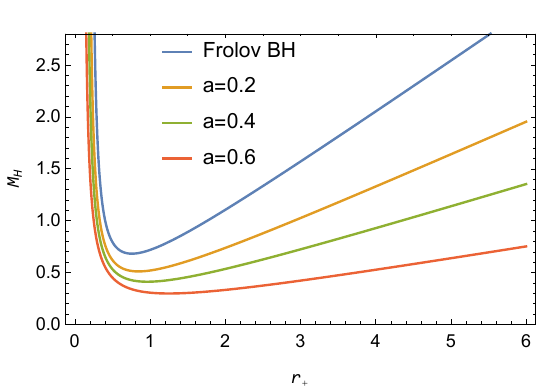}
\end{center}
\caption{The mass of Frolov BH with GM and CS showing the influence of the length scale parameter (left) and the CS parameter (right). Here $q=0.6$ and $\eta=0.4$.}\label{figT0}
\end{figure}

Figure \ref{figT0} illustrates how the BH parameters impact the mass function. The mass exhibits characteristic behavior: it initially decreases exponentially, reaching a minimum at a critical horizon radius, after which it increases monotonically with increasing horizon radius. This behavior is consistent with both the original Frolov BH and the RN BH. Our analysis focuses primarily on the parameters $\alpha$ and $a$, as the behavior of $q$ and $\eta$ is comparable to that of $\alpha$ and $a$, respectively.

Next, we determine the Hawking temperature based on the relationship between surface gravity and metric components in the semi-classical framework \cite{isz66}. The temperature of a BH is proportional to its surface gravity $\kappa$ at the event horizon through the relation $T = \kappa/2\pi$ \cite{isz67}. For the Frolov BH with GM and CS, the Hawking temperature takes the form:

\begin{equation}
T_{+}=\frac{q^2(4r^2_+\alpha^2(a+\eta^2-1)-r^4_+)-r_+^4(a+\eta^2-1)(3\alpha^2(a+\eta^2-1)+r^2_+)+\alpha^2(a+\eta^2-1)^2}{4 \pi \,r^3_+(r^4_++q^2\alpha^2)}.  \label{temp1}
\end{equation}

The temperature expression in Eq. (\ref{temp1}) reveals how the thermodynamic behavior of the BH is influenced by the combined effects of the Frolov parameter and topological defects \cite{isz68}. This modification has significant implications for the BH's evaporation process and final state, potentially avoiding the complete evaporation predicted for classical BHs \cite{isz69}. In specific limits, we recover known results: without GM and CS ($a=0=\eta$), Eq. (\ref{temp1}) reduces to the temperature of the original Frolov BH \cite{isz64}, and when $\alpha=0$, it reduces to the temperature of the RN BH as $T_+=\frac{r_{+}^{2}-q^2}{4\pi \,r_{+}^3}$.

\begin{figure}[ht!]
\begin{center}
\includegraphics[scale=0.9]{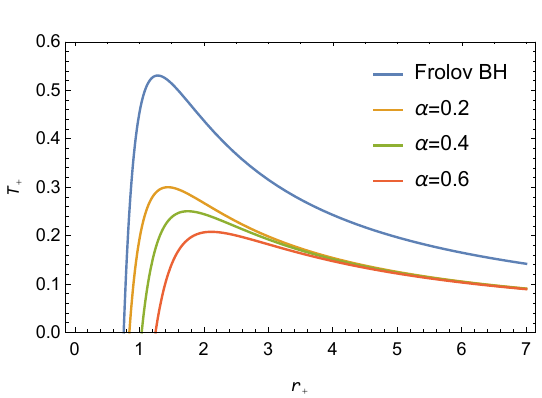}\quad
\includegraphics[scale=0.9]{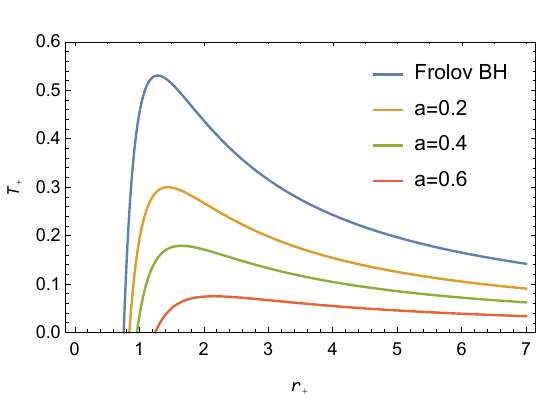}
\end{center}
\caption{The temperature of Frolov BH with GM and CS showing the influence of the length scale parameter (left) and the CS parameter (right). Here $q=0.6$ and $\eta=0.4$.}\label{figT1}
\end{figure}

Figure \ref{figT1} illustrates how the Hawking temperature $T_+$ varies with horizon radius and model parameters. The temperature initially rises with increasing horizon radius, reaches a peak, and then falls as the horizon radius increases further. The peak temperature decreases as $\alpha$ increases (with $a$, $\eta$, and $q$ held constant), and also decreases with increasing charge $q$. This behavior is consistent with both the original Frolov BH and the RN BH. As with the mass analysis, we focus primarily on the parameters $\alpha$ and $a$, since $q$ and $\eta$ exhibit behaviors similar to those of $\alpha$ and $a$, respectively.

To determine the entropy of the BH, we apply the first law of thermodynamics ($dM=T_{+}dS_{+}+\Phi\,dq$), which yields:

\begin{equation}
S_{+}=\int \frac{1}{T_{+}}\frac{\partial M}{\partial r_{+}}dr_{+}=\frac{\pi}{4}\left(r_{+}^2-\frac{\alpha^2(2q^2+\alpha^2(a+\eta^2-1)^2)}{r_{+}^2+\alpha^2(a+\eta^2-1)}-2\alpha^2(a+\eta^2-1)    \right)\log[r_{+}^2+\alpha^2(a+\eta^2-1)].
\label{entr2}
\end{equation}

The entropy expression in Eq. (\ref{entr2}) deviates significantly from the Bekenstein-Hawking area law that characterizes standard BH solutions \cite{isz70}. In classical BH thermodynamics, the entropy is proportional to the area of the event horizon, $S = A/4$. However, for the Frolov BH with GM and CS, the entropy contains additional terms that reflect the modified causal structure of the spacetime \cite{isz71}. This deviation has profound implications for information theory and the resolution of the information paradox \cite{isz72}.

To analyze the thermodynamic stability of the system, we calculate the specific heat capacity ($C_+$), which describes how the BH responds to temperature fluctuations. A positive value of specific heat indicates local stability (the system can absorb heat while maintaining a moderate temperature increase), whereas negative values imply local instability (temperature increases as mass decreases). The specific heat is determined using $C_{+}=\frac{dM}{dT_+}$, which gives:

\begin{equation}
    C_+=\frac{-2\pi\,r_{+}^4(r_{+}^4+2q^2\alpha^2)^2}{2q^4\alpha^2(r_{+}^2+\alpha^2 X)(r_{+}^2+3\alpha^2 X)-r_{+}^8X(r_+^2+9\alpha^2X)+q^2r_{+}^4(-3r_+^4+26r_+^2\alpha^2X+13\alpha^4X^2)}, \label{heatc1}
\end{equation}
where $X=a+\eta^2-1$.

The heat capacity expression in Eq. (\ref{heatc1}) reveals the complex thermodynamic structure of the Frolov BH with GM and CS, with multiple phase transitions possible depending on parameter values \cite{isz73}. The sign changes in the heat capacity indicate transitions between thermodynamically stable and unstable phases, which have significant implications for the BH's evolutionary path and final state \cite{isz74}. In specific limits, we recover known results: without GM and CS ($a=0=\eta$), Eq. (\ref{heatc1}) reduces to the heat capacity of the original Frolov BH \cite{isz64}, and when $\alpha=0$, it reduces to the specific heat capacity of the RN BH as $C_+=\frac{2\pi(3q^2-r_{+}^{2})}{r_{+}^4}$.

\begin{figure}[ht!]
\begin{center}
\includegraphics[scale=0.65]{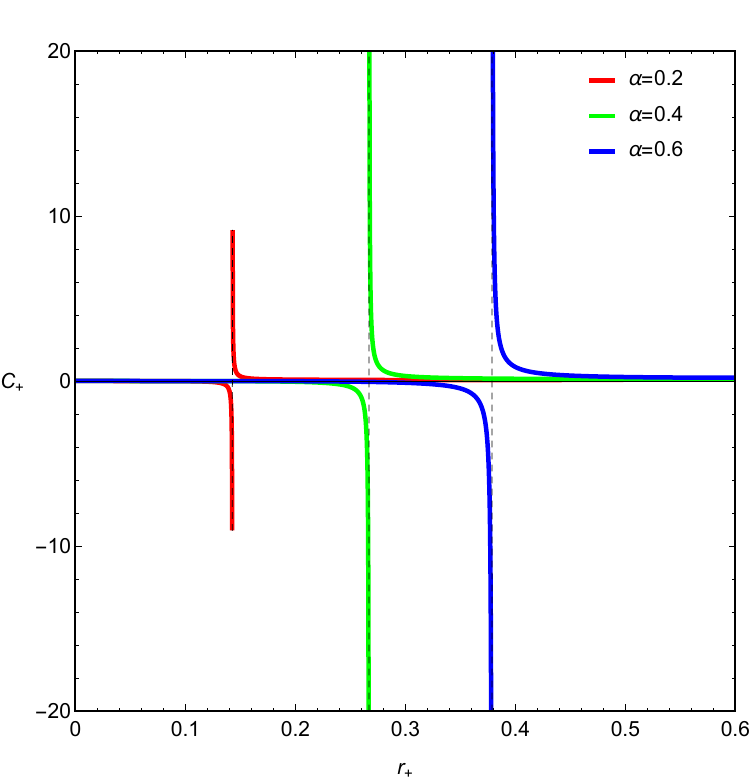}\quad
\includegraphics[scale=0.65]{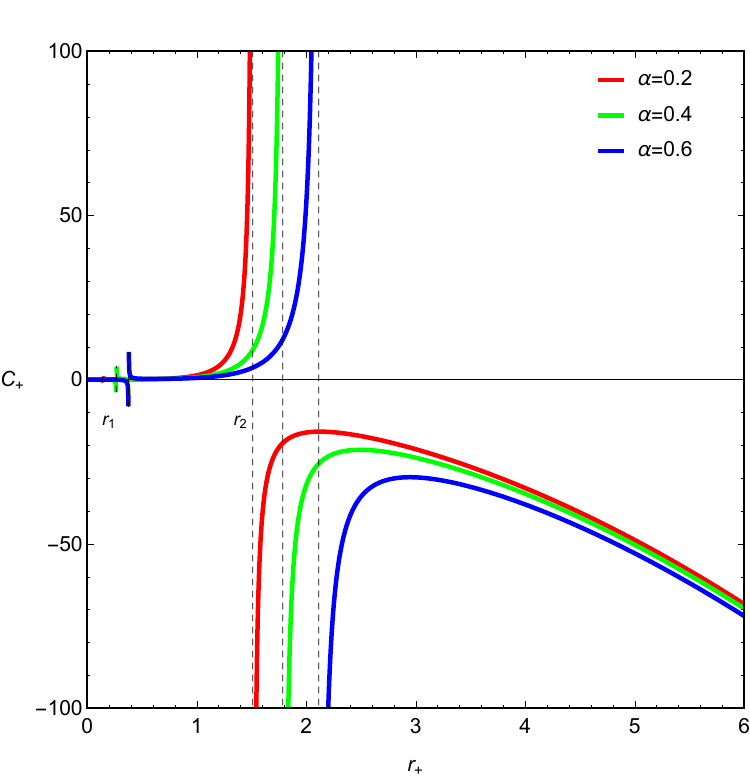}
\end{center}
\caption{The specific heat capacity of Frolov BH with GM and CS showing the influence of the length scale parameter for small values of horizon $r_+$ (left) and large values (right). Here $q=0.6$ and $\eta=0.4$.}\label{figT3}
\end{figure}

\begin{figure}[ht!]
\begin{center}
\includegraphics[scale=0.65]{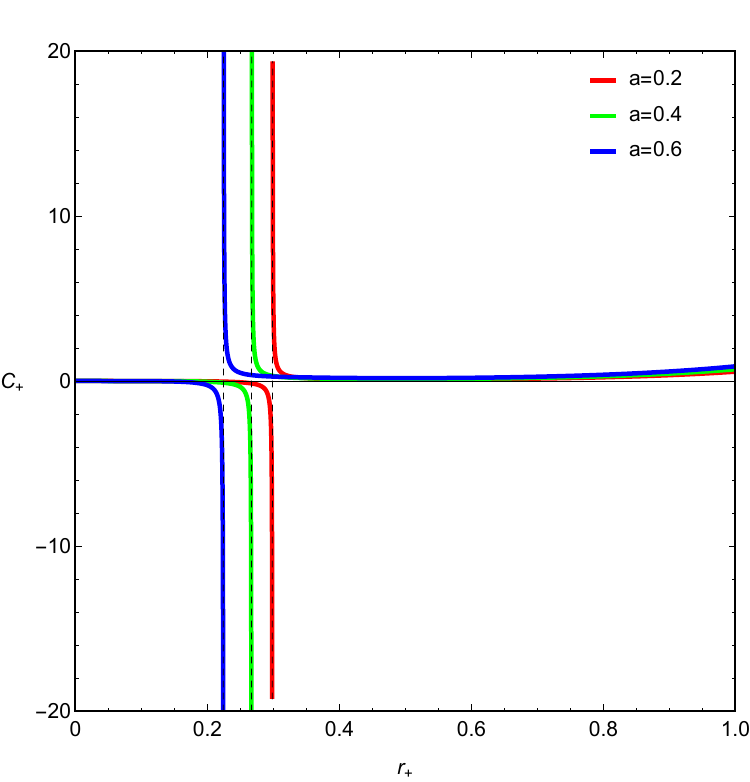}\quad
\includegraphics[scale=0.65]{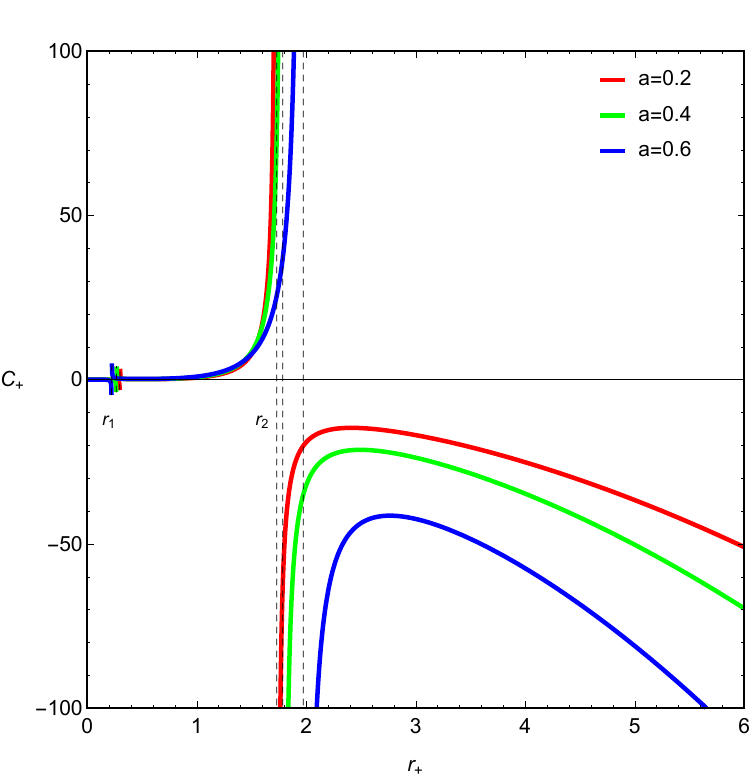}
\end{center}
\caption{The specific heat capacity of Frolov BH with GM and CS showing the influence of the CS parameter for small values of horizon $r_+$ (left) and large values (right). Here $q=0.6$ and $\eta=0.4$.}\label{figT4}
\end{figure}

Equation (\ref{heatc1}) for heat capacity is intricate and challenging to interpret, hence it is depicted in Figures \ref{figT3} and \ref{figT4}. Figure \ref{figT3} shows the heat capacity for various values of the length scale parameter $\alpha$ (with fixed values $a=0.2$, $q=0.6$, and $\eta=0.4$). The heat capacity diverges at two critical radii, $r_1$ and $r_2$, indicating a stable region for $r_1 < r_+ < r_2$ and an unstable region for $r_+ > r_2$. This behavior reveals a phase transition at $r_2$ from a smaller stable BH to a larger unstable BH. Figure \ref{figT4} shows similar behavior for various values of the CS parameter $a$ (with fixed values $\alpha=0.2$, $q=0.6$, and $\eta=0.4$).

The phase structure revealed by this heat capacity analysis has important implications for the BH's evaporation process and final state \cite{isz75}. The stable region between $r_1$ and $r_2$ suggests that the BH may reach a thermodynamic equilibrium state rather than evaporating completely, potentially resolving some of the paradoxes associated with the final stages of BH evaporation \cite{isz76}. Furthermore, the influence of the GM and CS parameters on the phase transitions indicates that topological defects can fundamentally alter the thermodynamic behavior of BHs, providing potential observational signatures of such exotic objects \cite{isz77}.

\section{Shadow OF FROLOV BH WITH GM AND CS} \label{sec05}

The detection of BH shadows by the EHT collaboration has opened a new observational window into strong-field gravity, making the theoretical investigation of BH shadows increasingly relevant for testing alternative gravity theories and exotic BH models \cite{isz03}. The BH shadow represents the apparent image of the photon capture region as seen by a distant observer, effectively marking the boundary between photons that escape to infinity and those captured by the BH \cite{isz28}. For modified BH solutions such as the Frolov BH with GM and CS, the shadow characteristics can serve as distinctive observational signatures that potentially distinguish these models from classical GR solutions \cite{isz81,isz82}.

The formation of a BH shadow is intimately connected to the existence of a photon sphere-a region where light rays can orbit the BH in unstable circular paths. For spherically symmetric spacetimes, the photon sphere corresponds to the critical radius where the effective potential for null geodesics reaches its maximum \cite{isz83}. The shadow radius, which determines the apparent size of the BH as seen by a distant observer, is then calculated based on the impact parameter of the critical null geodesics that define the boundary of the shadow \cite{isz84}.

This section examines how model parameters of Frolov BH with GM and CS affect the shadow radius. Spherically symmetric and static metrics with a photon sphere allow an observer to view an infinite number of images of a light source. We can determine the photon orbit radius using the following relation 
\begin{equation}
  r_{ph}\mathcal{F}'(r_{ph})=2\mathcal{F}(r_{ph}).
\end{equation}

This condition arises from analyzing the geodesic equations for null particles in the Frolov BH with GM and CS spacetime. By considering the Lagrangian for geodesic motion and applying the Euler-Lagrange equations, one obtains equations of motion that govern the trajectory of light rays \cite{isz86}. The condition for circular photon orbits is derived by setting the radial velocity and acceleration to zero, leading to the characteristic equation above that determines the radius of the photon sphere \cite{isz87}.

Considering the metric function in Eq. (\ref{bb2}), the equation for $r_{ph}$ becomes: 
\begin{equation}
3r_{ph}^7+(r_{ph}^4+(q^2+2r_{ph})\alpha^2)^2(a-\eta^2-1)-2q^2r_{ph}^3(\alpha^2+r_{ph}^3)=0. \label{eps1}
\end{equation} 

The complexity of Eq. (\ref{eps1}) reflects the intricate interplay between the Frolov parameter $\alpha$, the charge $q$, and the topological defect parameters $a$ and $\eta$ to determine the location of the photon sphere. Unlike the Schwarzschild BH, where the photon sphere occurs at a fixed radius $r_{ph} = 3M$, or the RN BH, where the photon sphere depends only on mass and charge, the Frolov BH with GM and CS exhibits a more complex dependence on multiple parameters \cite{isz88}. This complexity arises from the modified gravitational potential due to both the regular Frolov core and the topological defects, which alter the effective potential experienced by photons \cite{isz89}.

Since Eq. (\ref{eps1}) cannot be solved analytically, we employ numerical techniques to determine the photon orbit radii $r_{ph}$. Once we obtain these values, we calculate the shadow radii using: 
\begin{equation}
    R_s=\frac{r_{ph}}{\sqrt{\mathcal{F}(r_{ph})}}. \label{shadeq1}
\end{equation}

The shadow radius formula in Eq. (\ref{shadeq1}) represents the apparent size of the BH as seen by a distant observer in the equatorial plane. Physically, it corresponds to the critical impact parameter that separates the capture and scattering orbits for light rays approaching the BH \cite{isz90}. For observers at infinity, this critical impact parameter manifests itself as the radius of a dark circular region against a bright background, the BH shadow.

Tables \ref{taba3} and \ref{taba4} present numerical values of the photon sphere and shadow radii for various combinations of parameters of the Frolov BH with GM and CS. These tabulated results allow for a systematic analysis of how each parameter influences the shadow characteristics, providing quantitative predictions that could potentially be compared with observational data from instruments such as the EHT \cite{isz91}.

\begin{center}
\begin{tabular}{|c|c|c|c|c|c|c|}
 \hline   & \multicolumn{2}{|c|}{ $q=0.2$} & \multicolumn{2}{|c|}{ $0.4$} &  \multicolumn{2}{|c|}{ $0.6$} \\ \hline $\alpha$ & $r_{ph}$ & $R_{s}$ & $r_{ph}$ & $R_{s}$ & $r_{ph}$ & $R_{s}$  \\ \hline
$0.2$ & $6.78791$ & $17.7464$ & $6.70612$ & $17.5868$ & $6.56508$ & $17.3127$
\\ 
$0.4$ & $6.77736$ & $17.7327$ & $6.69504$ & $17.5724$ & $6.55301$ & $17.2972$
\\ 
$0.6$ & $6.75961$ & $17.7096$ & $6.67639$ & $17.5482$ & $6.53265$ & $17.2709$ \\ 
 
 \hline
\end{tabular}
\captionof{table}{Numerical results for the photon radius and shadow radius with various BH parameters, $\alpha$ and $q$. Here $a=0.4$ and $\eta=0.4$.} \label{taba3}
\end{center}

\begin{center}
\begin{tabular}{|c|c|c|c|c|c|c|}
 \hline   & \multicolumn{2}{|c|}{ $\eta=0.2$} & \multicolumn{2}{|c|}{ $0.4$} &  \multicolumn{2}{|c|}{ $0.6$} \\ \hline $a$ & $r_{ph}$ & $R_{s}$ & $r_{ph}$ & $R_{s}$ & $r_{ph}$ & $R_{s}$  \\ \hline
$0.2$ & $3.63485$ & $7.40957$ & $4.39661$ & $9.70029$ & $6.55301$ & $17.2972$
\\ 
$0.4$ & $5.0779$ & $11.9337$ & $6.55301$ & $17.2972$ & $12.2508$ & $43.5389$
\\ 
$0.6$ & $8.07556$ & $23.5079$ & $12.2508$ & $43.5389$ & $74.7591$ & $647.955$ \\ 
 
 \hline
\end{tabular}
\captionof{table}{Numerical results for the photon radius and shadow radius with various BH parameters, $a$ and $\eta$. Here $\alpha=0.4$ and $q=0.6$.} \label{taba4}
\end{center}

The shadow radius exhibits particular sensitivity to the parameters of the topological defects, as evidenced by the dramatic increase in shadow size with increasing values of $a$ and $\eta$ in Table \ref{taba4}. This enhanced sensitivity could provide a distinctive observational signature for the identification of BHs with topological defects, as the shadow size would appear significantly larger than expected for a classical BH of equivalent mass \cite{isz92}.

\begin{figure}
   \includegraphics[scale=0.65]{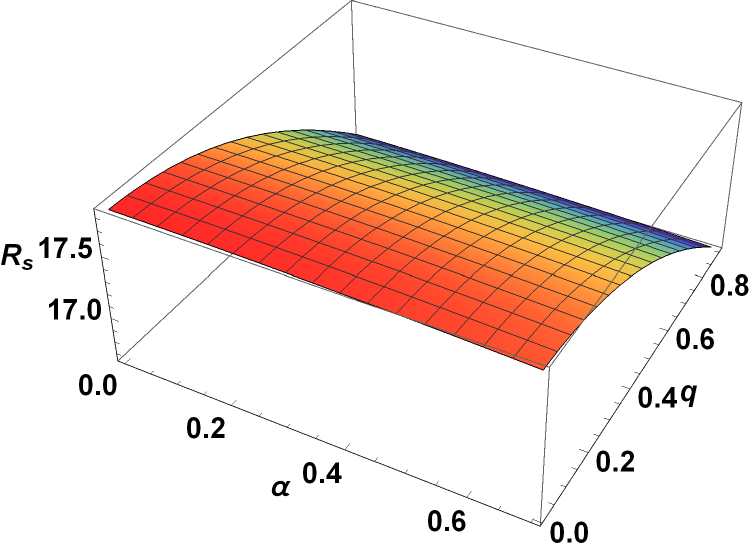}\quad
   \includegraphics[scale=0.65]{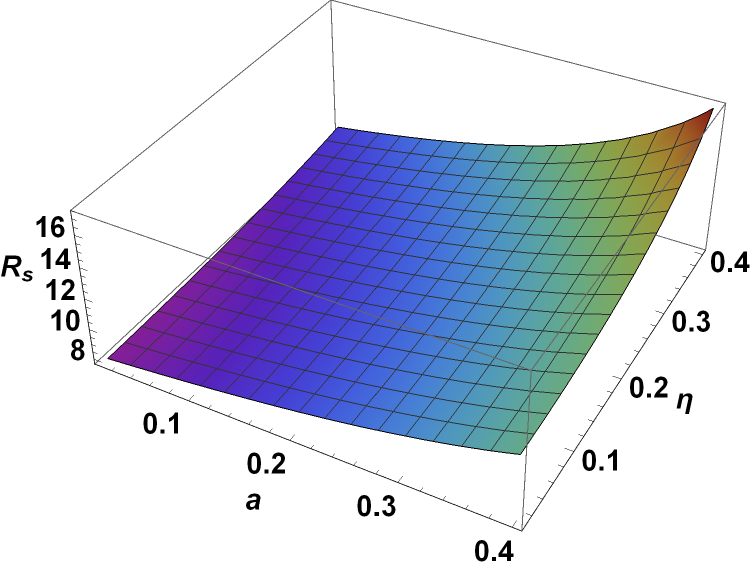}
    \caption{Variation of the shadow radius for different values of $\alpha$ and $q$ (left), and for different values of $\eta$ and $a$ (right).}
    \label{shad11}
\end{figure}

\begin{figure}
   \includegraphics[scale=0.7]{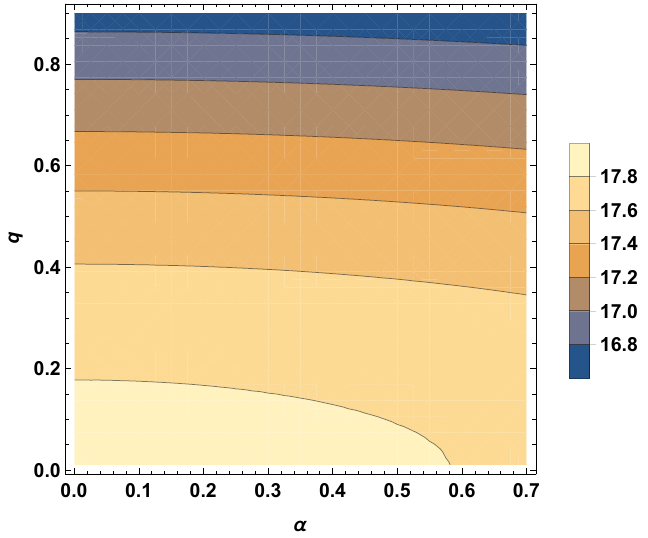}\quad
   \includegraphics[scale=0.7]{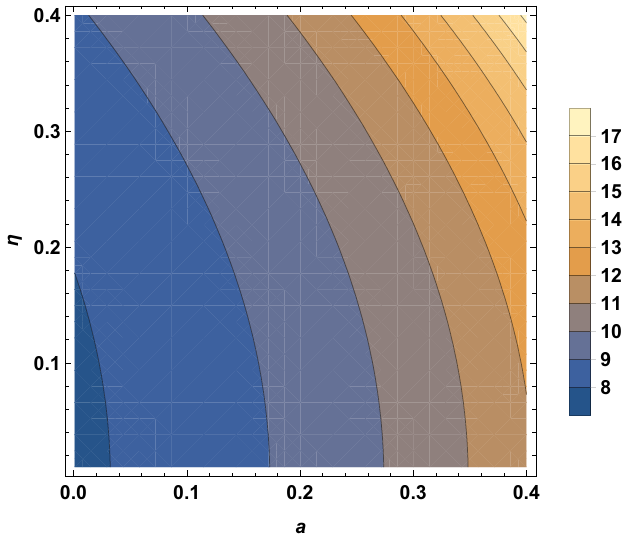}
    \caption{Contour plot of the shadow radius for different values of $\alpha$ and $q$ (left), and for different values of $\eta$ and $a$ (right).}
    \label{shad12}
\end{figure}

Tables \ref{taba3} and \ref{taba4} reveal an interesting pattern: Frolov BH parameters ($q$, $\alpha$) and topological defect parameters ($\eta$, $a$) affect the shadow radius in opposite ways. Increasing the Frolov parameters ($q$, $\alpha$) decreases the shadow radius, while increasing the topological defect parameters ($\eta$, $a$) increases it. Figures \ref{shad11} and \ref{shad12} illustrate these opposing tendencies using three-dimensional visualizations and contour plots, providing a comprehensive view of the dependencies of the parameters.

These opposing tendencies reveal the complex interplay between the Frolov regularization and topological defects in shaping the causal structure of spacetime. The decrease in shadow radius with increasing Frolov parameters ($q$, $\alpha$) stems from the regularization of the central singularity, which modifies the near-horizon geometry and effectively reduces the photon capture cross-section. Conversely, the increase in shadow radius with increasing topological defect parameters ($\eta$, $a$) reflects the additional gravitational effects introduced by the GM and CS, which enhance the light-bending capabilities of the spacetime and consequently enlarge the photon capture region.

The contour plots in Fig. \ref{shad12} offer valuable visualizations of these parameter dependencies, showing how the shadow radius varies across the parameter space. These visualizations could guide observational efforts by indicating which parameter regimes would produce the most distinctive shadow characteristics compared to classical BH solutions. The dramatic increase in shadow radius for large values of the topological defect parameters suggests that BHs with significant GM and CS components would exhibit shadows that are substantially larger than their classical counterparts, potentially providing a clear observational signature \cite{isz95}.

From an observational perspective, these results suggest that BHs with GM and CS could be distinguished from classical BHs through precise shadow measurements. Current and future observations by the EHT collaboration and other facilities could potentially constrain the presence and magnitude of such topological defects in astrophysical BHs by comparing measured shadow sizes with theoretical predictions across the parameter space \cite{isz96}. Additionally, the distinctive parameter dependencies identified in this analysis could help break degeneracies in the interpretation of shadow observations, aiding in the discrimination between different theoretical models \cite{isz97}.

\section{Scalar and EM Perturbations in Frolov BH geometry} \label{sec06}

Perturbation analysis serves as a powerful tool for probing the stability and spectral characteristics of BH spacetimes. It allows researchers to study how BHs respond to small external disturbances, offering deep insights into their dynamical behavior and potential observational signatures. Among the various types of perturbations-scalar, vector, and tensor-scalar field perturbations are particularly valuable due to their relative mathematical simplicity. Despite being simpler than gravitational (tensor) perturbations, scalar perturbations retain the ability to capture essential features of the BH's dynamical response and provide a useful approximation for understanding wave propagation and stability in curved spacetimes.

In the context of the Frolov BH, which incorporates both a regular core and non-trivial topological structures such as a GM and CS corrections, the analysis of scalar field perturbations becomes especially meaningful. These perturbations help elucidate how the presence of these dual topological defects influences the quasinormal modes (QNMs)-the characteristic oscillation frequencies of BHs-and ultimately affects the overall stability of the spacetime \cite{isz95}. The combined effects of the regular geometry near the core and the asymptotically non-flat structure induced by the GM and CS terms result in a modified perturbative spectrum, which could, in principle, be probed through gravitational wave observations.

Understanding scalar perturbations is also crucial in a broader context, as they have been widely employed to investigate the stability of a range of BH solutions in general relativity and beyond. For example, detailed studies have been carried out for Schwarzschild, Kerr, and RN BHs, where scalar field perturbations provide valuable information about the propagation of matter and fields in curved spacetime and the response of the geometry to such perturbations \cite{isz96}. These investigations have been extended to BHs in alternative theories of gravity, offering a comparative perspective on stability across different gravitational frameworks. Overall, the study of scalar perturbations in BH spacetimes not only enhances our theoretical understanding of BH dynamics but also contributes to the ongoing efforts in gravitational wave astronomy, where such perturbative signatures may eventually become key observational indicators of exotic spacetime structures and new physics.

In this section, we investigate scalar field perturbations in the background of the Frolov BH incorporating a global monopole and cosmic string term. Our analysis begins with the derivation of the massless Klein-Gordon equation, which governs the evolution of a scalar field in this spacetime geometry. The study of perturbations typically starts with identifying the appropriate field equation. For a massless scalar field, the dynamics are described by the Klein–Gordon equation in curved spacetime, expressed in its covariant form as \cite{CJPHY,NPB,AHEP4,EPJC}:
\begin{equation}
\frac{1}{\sqrt{-g}}\,\partial_{\mu}\left[\left(\sqrt{-g}\,g^{\mu\nu}\,\partial_{\nu}\right)\,\Psi\right]=0,\label{ff1}    
\end{equation}
where $\Psi$ is the wave function of the scalar field, $g_{\mu\nu}$ is the covariant metric tensor, $g=\det(g_{\mu\nu})$ is the determinant of the metric tensor, $g^{\mu\nu}$ is the contrvariant form of the metric tensor, and $\partial_{\mu}$ is the partial derivative with respect to the coordinate systems.

Similarly, for EM perturbations, the dynamics are governed by Maxwell's equations in curved spacetime:
\begin{equation}
\frac{1}{\sqrt{-g}}\partial _{\mu }\left[ F_{\alpha \beta }g^{\alpha \nu
}g^{\beta \mu }\sqrt{-g}\,\Psi\right] =0,  \label{eq22}
\end{equation}
where $F_{\alpha \beta }=\partial _{\alpha }A_{\beta }-\partial _{\beta}A_{\nu}$ is the EM tensor.

To solve these equations efficiently, we employ the tortoise coordinate transformation, which maps the semi-infinite domain $[r_+, \infty)$ to the entire real line $(-\infty, \infty)$, facilitating the analysis of wave propagation near the horizon and at spatial infinity. The tortoise coordinate is defined by:
\begin{eqnarray}
    dr_*=\frac{dr}{\mathcal{F}(r)}\label{ff2}
\end{eqnarray}

Applying this coordinate transformation to the line-element Eq. (\ref{bb1}) results in:
\begin{equation}
    ds^2=\mathcal{F}(r_*)\,\left(-dt^2+dr^2_{*}\right)+\mathcal{D}^2(r_*)\,\left(d\theta^2+\sin^2 \theta\,d\phi^2\right),\label{ff3}
\end{equation}
where $\mathcal{F}(r_*)$ and $\mathcal{D}(r_*)$ are the metric functions expressed in terms of the tortoise coordinate $r_*$. 

To analyze perturbations, we employ the separation of variables technique by expressing the scalar field in the form:
\begin{equation}
    \Psi(t, r_{*},\theta, \phi)=\exp(i\,\omega\,t)\,Y^{m}_{\ell} (\theta,\phi)\,\frac{\psi(r_*)}{r_{*}},\label{ff4}
\end{equation}
where $\omega$ is the (possibly complex) temporal frequency, $\psi(r)$ is the radial part of the scalar field, and $Y^{m}_{\ell} (\theta,\phi)$ are the spherical harmonics characterized by the angular momentum quantum numbers $\ell$ and $m$.

With this ansatz, the wave equations (\ref{ff1}) and (\ref{eq22}) reduce to a one-dimensional Schrödinger-like equation:
\begin{equation}
    \frac{\partial^2 \psi(r_*)}{\partial r^2_{*}}+\left(\omega^2-V_\text{scalar}\right)\,\psi(r_*)=0,\label{ff5}
\end{equation}
where $V$ represents the effective potential that encapsulates the influence of the background spacetime on the propagation of the perturbation. For scalar perturbations, this effective potential takes the form:
\begin{eqnarray}
V_\text{scalar}(r)&=&\left(\frac{\ell\,(\ell+1)}{r^2}+\frac{\mathcal{F}'(r)}{r}\right)\,\mathcal{F}(r)\nonumber\\
&=&\left(1-a-8\,\pi\,\eta^2-\frac{(2\,Mr-q^2)r^2}{r^4+(2\,Mr+q^2)\alpha^2}\right)\left(\frac{\ell\,(\ell+1)}{r^2}+\frac{2r^4(Mr-q^2)+2(q^4-2Mq^2r-4M^2r^2)\alpha^2}{\left(r^4+(2\,Mr+q^2)\alpha^2 \right)^2}\right).\label{ff6}
\end{eqnarray}

Similarly, for the EM perturbations (vector field), the effective potential is given by
\begin{equation}
V_\text{EM}(r)=\frac{\ell\,(\ell+1)}{r^2}\mathcal{F}(r)= \frac{\ell\,(\ell+1)}{r^2} \left(1-a-8\,\pi\,\eta^2-\frac{(2\,Mr-q^2)r^2}{r^4+(2\,Mr+q^2)\alpha^2}\right).\label{ff7}
\end{equation}

The effective potentials for scalar and vector perturbations given in Eqs. (\ref{ff6}) and (\ref{ff7}) shows that various factors involved in the spacetime geometry influence these potentials, and thus, gets modification compared to the standard Schwarzschild or RN-BH solution. These factors include the length scale parameter $\alpha$, and the dual topological defects parameters ($a$, $\eta$). Additionally, the BH mass $M$ and electric charge $q$, and the multipole number $\ell$ alters these potentials. The structure of these potentials determines the characteristic oscillation modes-known as QNMs-that dominate the late-time response of the BH to external perturbations \cite{isz98}.

\begin{figure}[H]
   \includegraphics[scale=0.9]{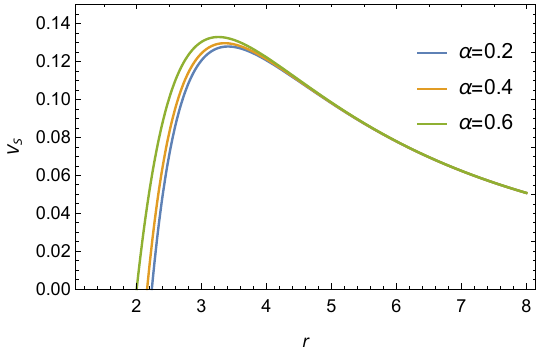}\quad
\includegraphics[scale=0.9]{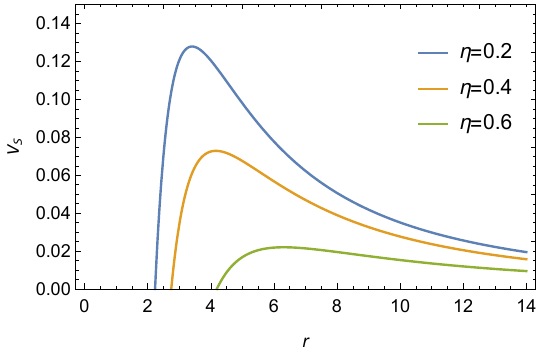}
    \caption{ The plot of the scalar potential is shown for different combinations of the parameter. In the left panel, for different values of $\alpha$. In the right panel, for different values of $\eta$. Here, $a=0.2$, $q=0.8$, and $l=2$.}
    \label{potent1}
\end{figure}

\begin{figure}[H]
   \includegraphics[scale=0.9]{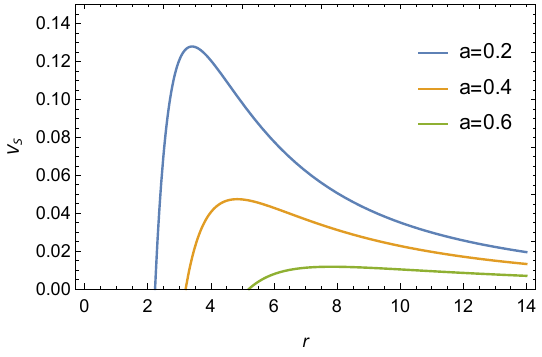}\quad
\includegraphics[scale=0.9]{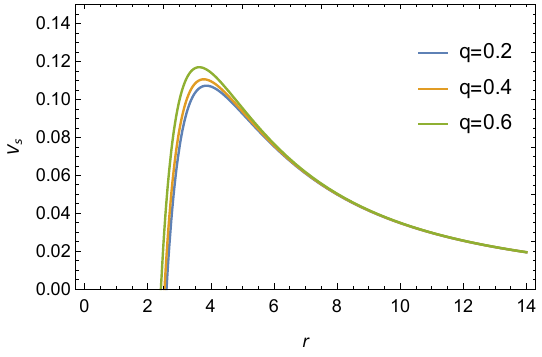}
    \caption{ The plot of the scalar potential is shown for different combinations of the parameter. In the left panel, for different values of $a$. In the right panel, for different values of $q$. Here, $\alpha=0.2$, $\eta=0.2$, and $l=2$.}
    \label{potent2}
\end{figure}

It is evident that Eqs. (\ref{ff6}) and (\ref{ff7}) are influenced by several factors, including the CS parameter $a$, the GM parameter $\eta$, and the length scale parameter $\alpha$. In Figs. \ref{potent1} and \ref{potent2}, we present the radial evolution of the scalar perturbative potential for different values of BH parameters. The figures show how the parameters $\alpha$, $\eta$, $a$, and $q$ affect the perturbative potentials. 
\begin{itemize}
     \item The potential appears to be somewhat insensitive to the parameters $\alpha$ and $q$ when compared to the parameters  $\eta$ and $a$.
     \item As the length scale parameter $\alpha$ increases, the potential peak falls, and the charge parameter $q$ follows a similar trend.
     \item As the GM parameter $\eta$ increases, so does the potential, for $a$, following a similar trend.
 \end{itemize}
 
The shape of the effective potential provides valuable insights into the stability and spectral characteristics of the BH. A positive definite potential with a single peak, as observed in Figs. \ref{potent1} and \ref{potent2}, indicates stability against scalar perturbations, as it prevents the existence of bound states with negative energy that could trigger instabilities. The height and width of the potential barrier determine the oscillation frequency and damping rate of the QNMs, with higher barriers typically corresponding to higher frequencies and longer damping times.

\begin{figure}[ht!]
    \centering
    \includegraphics[width=18cm]{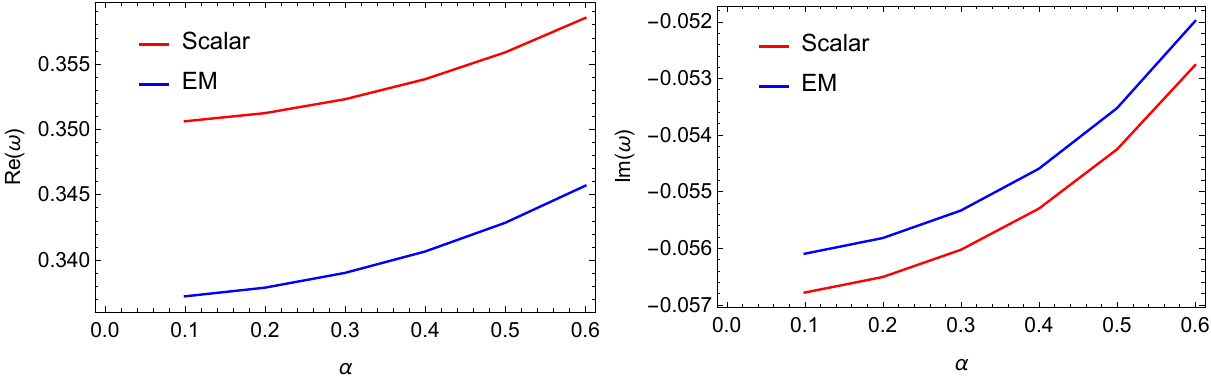}
    \caption{ Variation of amplitude and damping of QNMs with respect to the length scale parameter $\alpha$  for scalar and EM perturbations.}
    \label{realA}
\end{figure}

\begin{figure}[ht!]
    \centering
    \includegraphics[width=18cm]{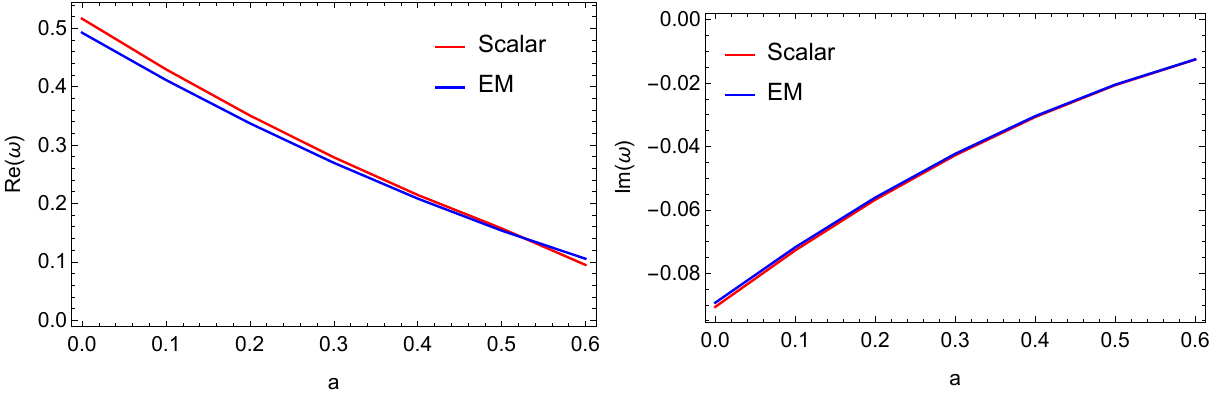}
    \caption{ Variation of amplitude and damping of QNMs with respect to the CS  parameter $a$ for scalar and EM perturbations.}
    \label{realAa}
\end{figure}

\begin{figure}[ht!]
    \centering
    \includegraphics[width=18cm]{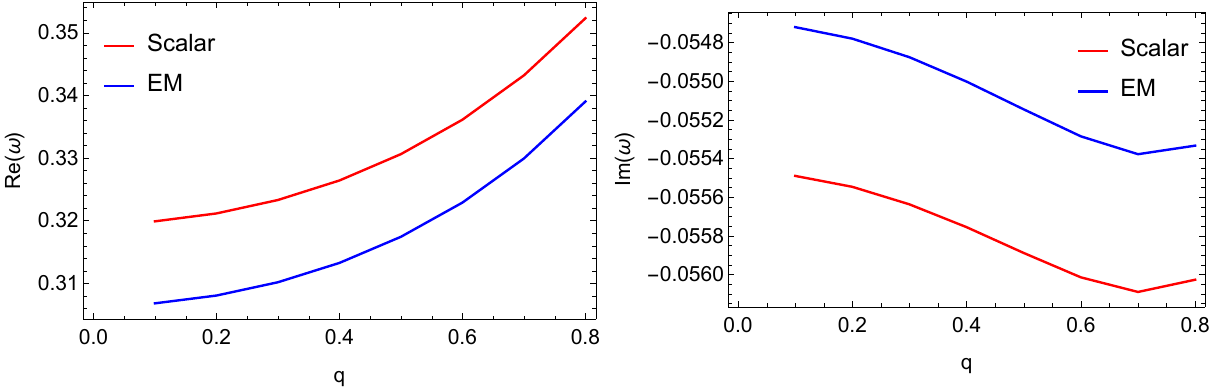}
    \caption{ Variation of amplitude and damping of QNMs with respect to the charge  parameter $q$ for scalar and EM perturbations.}
    \label{realQ}
\end{figure}

The WKB approximation method is often used to calculate QNMs. It was first introduced by Iyer \cite{d3} and later expanded to higher levels by Konoplya \cite{d4}. The WKB technique is effective for low overtone values $n$, especially for $n <\ell$. Using the perturbative potentials, we numerically calculate the quasinormal frequencies for scalar and EM perturbations using the 6th order WKB approximation. The results of the calculated QNMs are presented in Tables \ref{taba13}-\ref{taba16}, which show the dependence of the QNM frequencies on the BH parameters $\alpha$, $\eta$, $a$ and $q$. Figures \ref{realA}- \ref{realQ} summarize the results from tables \ref{taba13}-\ref{taba16}. 

\begin{center}
\begin{tabular}{|c|c|c|}
 \hline 
 \multicolumn{3}{|c|}{ $q=0.8$, $\alpha=0.2$, $n=0$, $\ell=2$, $a=0.2$}
\\ \hline $\eta $ & $Scalar$ & $EM$ \\ \hline
$0$ & $0.362932-0.0627145i$ & $0.368547-0.0610567i$ \\ 
$0.1$ & $0.355801-0.0611452i$ & $0.361048-0.0596077i$ \\ 
$0.2$ & $0.330296-0.0550463i$ & $0.339009-0.055334i$ \\ 
$0.3$ & $0.300809-0.0493128i$ & $0.303753-0.0484837i$ \\ 
$0.4$ & $0.255723-0.0400121i$ & $0.257378-0.0395421i$ \\ 
$0.5$ & $0.202005-0.029494i$ & $0.202678-0.029279i$
\\ 
 \hline
\end{tabular}
\captionof{table}{Variation of amplitude and damping of QNMs with respect to the energy scale of the spontaneous symmetry
breaking parameter.} \label{taba13}
\end{center} 

\begin{center}
\begin{tabular}{|c|c|c|}
 \hline 
 \multicolumn{3}{|c|}{ $q=0.8$, $n=0$, $\ell=2$, $\alpha=0.2$, $\eta=0.2$}
\\ \hline $a$ & $Scalar$ & $EM$ \\ 
\hline
$0$ & $0.516159-0.0906914i$ & $0.492135-0.0893779i$ \\ 
$0.1$ & $0.429441-0.0728478i$ & $0.411179-0.071879i$ \\ 
$0.2$ & $0.350406-0.0568682i$ & $0.337001-0.0561831i$ \\ 
$0.3$ & $0.278756-0.0428282i$ & $0.269344-0.0423692i$ \\ 
$0.4$ & $0.214322-0.0307679i$ & $0.208089-0.0304816i$ \\ 
$0.5$ & $0.157065-0.0207048i$ & $0.153261-0.0205431i$ \\ 
$0.6$ & $0.107105-0.0126417i$ & $0.105049-0.0125628i$
\\ 
 \hline
\end{tabular}
\captionof{table}{Variation of amplitude and damping of QNMs with respect to CS parameter.} \label{taba14}
\end{center} 
\begin{center}
\begin{tabular}{|c|c|c|}
 \hline 
 \multicolumn{3}{|c|}{  $\eta=0.2$, $n=0$, $\ell=2$, $a=0.2$, $q=0.8$}
\\ \hline $\alpha$ & $Scalar$ & EM \\ \hline
$0.1$ & $0.350604-0.0567825i$ & $0.337212-0.0560965i$ \\ 
$0.2$ & $0.35123-0.0565079i$ & $0.337877-0.0558201i$ \\ 
$0.3$ & $0.352294-0.0560271i$ & $0.339009-0.055334i$ \\ 
$0.4$ & $0.353828-0.0552993i$ & $0.340645-0.054595i$ \\ 
$0.5$ & $0.355881-0.0542537i$ & $0.342843-0.0535259i$ \\ 
$0.6$ & $0.358512-0.0527671i$ & $0.345676-0.0519899i$
\\ 
 \hline
\end{tabular}
\captionof{table}{Variation of amplitude and damping of QNMs with respect to length scale parameter.} \label{taba15}
\end{center} 
\begin{center}
\begin{tabular}{|c|c|c|}
 \hline 
 \multicolumn{3}{|c|}{ $a=0.2$, $n=0$, $\ell=2$, $\alpha=0.2$, $\eta=0.2$}
\\ \hline $q$ & $Scalar$ & $EM$ \\ 
\hline
$0.1$ & $0.319889-0.0554904i$ & $0.306802-0.0547214i$ \\ 
$0.2$ & $0.321148-0.0555469i$ & $0.308045-0.054781i$ \\ 
$0.3$ & $0.323294-0.0556375i$ & $0.310168-0.054877i$ \\ 
$0.4$ & $0.326408-0.0557555i$ & $0.313252-0.055003i$ \\ 
$0.5$ & $0.330616-0.0558888i$ & $0.317423-0.0551465i$ \\ 
$0.6$ & $0.33611-0.0560148i$ & $0.322878-0.055286i$ \\ 
$0.7$ & $0.343183-0.0560903i$ & $0.329915-0.0553782i$ \\ 
$0.8$ & $0.352294-0.0560271i$ & $0.339009-0.055334i$
\\ 
 \hline
\end{tabular}
\captionof{table}{Variation of amplitude and damping of QNMs with respect to charge parameter.} \label{taba16}
\end{center} 

The QNM frequencies presented in Tables \ref{taba13}-\ref{taba16} consist of a real part, representing the oscillation frequency, and an imaginary part, representing the damping rate. The negative imaginary component indicates stability of the perturbations, as it ensures that the oscillations decay exponentially with time rather than growing unboundedly. This confirms the stability of the Frolov BH with GM and CS against scalar and EM perturbations, at least within the parameter ranges considered in this study.

The real part of the QNM frequency, $Re(\omega)$, which specifies the oscillation frequency of the perturbations, decreases as the GM parameter $\eta$ increases. This tendency implies that the oscillation frequency of the perturbations gradually decreases, resulting in a decreased response of the BH spacetime. In contrast, when $\eta$ increases, the imaginary part $Im(\omega)$, which represents the dissipative rate of the oscillations, becomes less negative. This suggests a slower dissipation rate of the disturbances, implying that the oscillations will persist for a longer period of time. A similar pattern is observed for the CS parameter $a$.

The study also examines the QNM frequencies in relation to the length scale parameter $\alpha$ and charge $q$. As $\alpha$ increases, $Re(\omega)$ rises, indicating more stable oscillations at higher levels. As $\alpha$ increases, $Im(\omega)$ decreases, perturbations dissipate more slowly. The charge $q$ has the opposite effect on QNM frequencies compared to $\alpha$. As $q$ increases, so does $Re(\omega)$, indicating quicker oscillations. As $q$ increases, $Im(\omega)$ becomes less negative, indicating slower decay of disturbances. The EM field associated with the charge parameter "tightens" the effective potential around the BH, enhancing the oscillatory response and duration of the oscillations.

\begin{figure}[ht!]
    \centering
    \includegraphics[width=18cm]{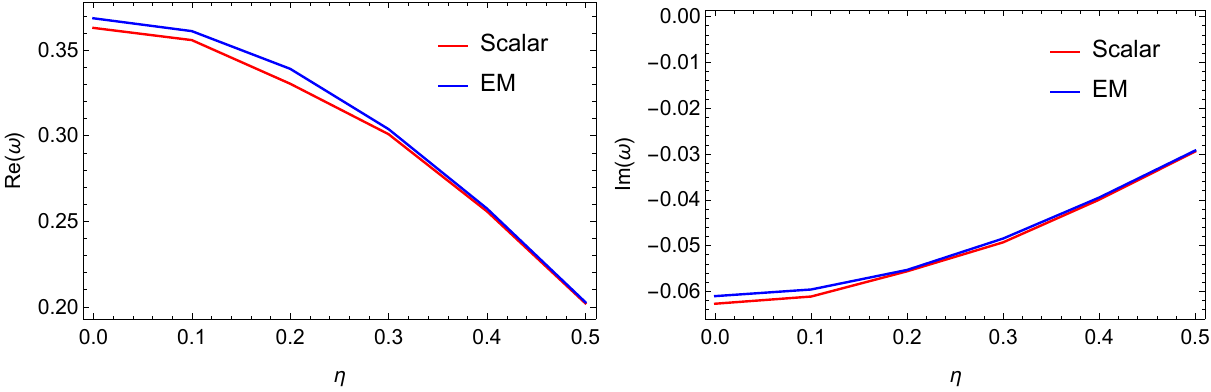}
    \caption{ Variation of amplitude and damping of QNMs with respect to the GM   parameter $\eta$ for scalar and EM perturbations.}
    \label{realT}
\end{figure}

The trends observed in the QNM spectra have significant physical implications. The slower damping rates associated with increasing topological defect parameters suggest that BHs with substantial GM and CS components would exhibit longer-lived ringdown signals following perturbation events such as binary mergers. This extended ringdown phase could potentially provide a distinctive observational signature in gravitational wave data, offering a means to identify or constrain the presence of topological defects in astrophysical BHs \cite{isz99}. Furthermore, the parameter-dependent behavior of the QNM frequencies reveals the intricate interplay between the regular core structure, EM field, and topological defects in shaping the spectral characteristics of the BH.

\section{Conclusion} \label{sec07}

In this work, we have presented a comprehensive investigation of the Frolov BH with GM and CS. We explored the combined effects of a non-singular BH core, as introduced by Frolov, and the modifications induced by topological defects. Our study systematically analyzed the geometric, thermodynamic, and perturbative properties of the composite spacetime, providing new insights into how the parameters $\alpha$, $q$, $\eta$, and $a$ interact to shape the BH characteristics.

We began by deriving the spacetime metric given by Eq. (\ref{bb1}) and the associated metric function $\mathcal{F}(r)$ in Eq. (\ref{bb2}). This formulation encapsulated both the regularizing feature introduced by the Frolov parameter $\alpha$ and the contributions of the GM and CS, represented by the parameters $\eta$ and $a$, respectively. Through numerical methods, we demonstrated that the lapse function and the corresponding horizon structure were significantly modified compared to standard BHs. Figures \ref{lapse1}, \ref{Horiz1}, and \ref{Horiz2} illustrated how the inner and outer horizons were affected by variations in $\alpha$, $q$, $a$, and $\eta$, confirming that the presence of topological defects induced notable deviations from the classical RN solution.

In our thermodynamic analysis, we derived the mass function in Eq. (\ref{mass1}) by imposing the horizon condition $\mathcal{F}(r_+)=0$, which allowed us to express the BH mass $M_H$ as a function of the horizon radius $r_+$ and the model parameters. Figure \ref{figT0} demonstrated that the mass function reached a minimum at a critical horizon radius before increasing monotonically-a behavior that paralleled the trends observed in both the original Frolov and RN BHs. This result not only validated our approach but also highlighted the intricate balance between the regularization parameter and the effects of the topological defects.

The Hawking temperature, derived in Eq. (\ref{temp1}), was shown to depend sensitively on the horizon radius as well as on $\alpha$, $q$, $a$, and $\eta$. Figure \ref{figT1} depicted a temperature profile that increased with $r_+$ until reaching a peak, followed by a subsequent decline. This peak shifted with variations in $\alpha$ and $q$, indicating that the interplay between the EM charge and the regularization parameter plays a crucial role in dictating the BH's evaporation dynamics. In the appropriate limits, our temperature expression reduced to that of the original Frolov BH or the standard RN solution, thus providing a consistency check for our model.

We computed the entropy via the first law of thermodynamics, leading to the expression in Eq. (\ref{entr2}). Unlike the conventional Bekenstein-Hawking area law, the entropy here included logarithmic corrections and additional terms arising from the modified causal structure due to the presence of the GM and CS. This deviation suggested that the information content of the BH is influenced by the topological defects, potentially offering a resolution to aspects of the information paradox \cite{isz98x,isz98xx}.

The analysis of the specific heat capacity, given by Eq. (\ref{heatc1}), revealed further subtleties in the thermodynamic behavior. Figures \ref{figT3} and \ref{figT4} clearly showed that the specific heat diverged at critical horizon radii, signaling phase transitions between thermodynamically stable and unstable regions. In particular, we observed that the BH exhibited a stable phase within a certain range of $r_+$ (between $r_1$ and $r_2$), beyond which it became unstable. These phase transitions were markedly influenced by the parameters $\alpha$, $a$, and $\eta$, implying that the topological defects not only alter the gravitational field but also have profound implications for the thermal stability and evolution of the BH \cite{isz99x}.

Our study also extended to the observational signatures of the BH through the analysis of its shadow. By determining the photon sphere via the condition $r_{ph}\mathcal{F}'(r_{ph}) = 2\mathcal{F}(r_{ph})$ and computing the shadow radius using Eq. (\ref{shadeq1}), we were able to generate numerical predictions for the apparent size of the BH. Tables \ref{taba3} and \ref{taba4} provided detailed numerical values of the photon sphere and shadow radii under different parameter configurations, while Figures \ref{shad11} and \ref{shad12} offered clear visualizations of how the shadow radius varied with $\alpha$, $q$, $a$, and $\eta$. Notably, our results demonstrated that while an increase in the Frolov parameters $\alpha$ and $q$ tended to decrease the shadow radius, the incorporation of topological defects (i.e., increasing $\eta$ and $a$) led to a significant enlargement of the shadow. This opposing trend could serve as a distinctive observational signature to differentiate between classical and modified BH models.

Finally, we thoroughly explored the dynamics of scalar and EM perturbations around the composite BH spacetime. By solving the massless Klein-Gordon and Maxwell equations, we derived and analyzed the effective potentials for scalar and EM perturbations both analytically and numerically. Our findings indicated that the height and width of the perturbative potentials were significantly sensitive to variations in the GM and CS parameters, whereas the impacts of the Frolov length scale $\alpha$ and electric charge $q$ were comparatively mild, as clearly shown in Figures \ref{potent1} and \ref{potent2}. Utilizing the 6th order WKB approximation method, we computed the QNMs for both scalar and EM perturbations. Our comprehensive analysis (see Tables \ref{taba13}-\ref{taba16} and Figures \ref{realT}-\ref{realQ}) revealed that increasing the GM and CS parameters led to notable decreases in both the oscillation frequencies and the decay rates of the perturbations, reflecting slower damping and prolonged ringdown phases. Importantly, this behavior was consistent across both scalar and EM perturbations, reinforcing the overall stability of the Frolov BH solutions modified by GM and CS. These extended ringdown signatures in gravitational wave observations could serve as distinctive indicators for the presence of topological defects in astrophysical BHs \cite{isz100x}.

In summary, our work successfully combined analytical and numerical methods to explore the multifaceted properties of the Frolov BH modified by GM and CS defects. We demonstrated that the inclusion of these topological defects significantly altered the horizon structure, thermodynamic behavior, shadow characteristics, and perturbative dynamics of the BH. Our findings not only corroborated known limits-reducing to standard Frolov or RN BHs under appropriate conditions-but also revealed novel features that could be probed in future observational campaigns.

Looking forward, we plan to extend this study in several directions. First, we will consider additional perturbative analyses, including EM and gravitational perturbations, which are expected to provide further insights into the dynamical stability of these exotic objects. Second, we aim to explore rotating configurations of the Frolov BH with GM and CS, as rotation is an essential feature of astrophysical BHs \cite{isz101x}. Finally, we plan to employ advanced numerical techniques and higher-order corrections to refine our QNM calculations and to simulate more realistic scenarios that could be compared with future gravitational wave observations \cite{isz102x}.

\section*{Acknowledgments}

F.A. expresses gratitude to the Inter University Centre for Astronomy and Astrophysics (IUCAA), Pune, India, for a visiting associateship. \.{I}.~S. acknowledges academic and/or financial support from EMU, T\"{U}B\.{I}TAK, ANKOS, and SCOAP3, along with networking support from COST Actions CA22113, CA21106, and CA23130.

\section*{Data Availability Statement}

There are no new data associated with this article.


\begin{thebibliography}{1}

\bibitem{isz01} B.~P.~Abbott et al. [LIGO Scientific and Virgo Collaborations], Phys.\ Rev.\ Lett.\  \textbf{116}, 061102 (2016).

\bibitem{isz02} B.~P.~Abbott et al. [LIGO Scientific and Virgo Collaborations], Phys.\ Rev.\ X \textbf{9}, 031040 (2019).

\bibitem{isz03} K.~Akiyama et al. [Event Horizon Telescope Collaboration], Astrophys.\ J.\ Lett.\  \textbf{875}, L1 (2019).

\bibitem{isz05} W.~Israel, Phys.\ Rev.\  \textbf{164}, 1776 (1967).

\bibitem{isz06} B.~Carter, Phys.\ Rev.\ Lett.\  \textbf{26}, 331 (1971).

\bibitem{frolov} V.~P.~Frolov, Phys.\ Rev.\ D \textbf{94}, 104056 (2016).

\bibitem{isz07} L.~Modesto and L.~Rachwał, Int.\ J.\ Mod.\ Phys.\ D \textbf{26}, 1730020 (2017).

\bibitem{isz08} E.~Ayon-Beato and A.~Garcia, Phys.\ Rev.\ Lett.\  \textbf{80}, 5056 (1998).

\bibitem{isz09} S.~A.~Hayward, Phys.\ Rev.\ Lett.\  \textbf{96}, 031103 (2006).

\bibitem{isz10} M.~Barriola and A.~Vilenkin, Phys.\ Rev.\ Lett.\  \textbf{63}, 341 (1989).

\bibitem{isz11} D.~Harari and C.~Lousto, Phys.\ Rev.\ D \textbf{42}, 2626 (1990).

\bibitem{isz12} N.~Dadhich, K.~Narayan and U.~A.~Yajnik, Pramana \textbf{50}, 307 (1998).

\bibitem{isz13} A.~Vilenkin, Phys.\ Rept.\  \textbf{121}, 263 (1985).

\bibitem{isz14} A.~Vilenkin and E.~P.~S.~Shellard, \textit{Cosmic Strings and Other Topological Defects} (Cambridge University Press, Cambridge (UK), 2000).

\bibitem{isz15} M.~B.~Hindmarsh and T.~W.~B.~Kibble, Rept.\ Prog.\ Phys.\  \textbf{58}, 477 (1995).

\bibitem{isz16} M.~Barriola and A.~Vilenkin, Phys. Rev. Lett. \textbf{63}, 341 (1989).

\bibitem{isz17} K.~A.~Bronnikov, B.~E.~Meierovich and E.~R.~Podolyak, J. Exp. Theor. Phys. \textbf{95}, 392 (2002).

\bibitem{isz18} H.~Tan, J.~Yang, J.~Zhang and T.~He, Chin.\ Phys.\ B \textbf{27}, 030401 (2018).

\bibitem{isz19} M.~Aryal, L.~H.~Ford and A.~Vilenkin, Phys.\ Rev.\ D \textbf{34}, 2263 (1986).

\bibitem{isz20} A.~Achucarro, R.~Gregory and K.~Kuijken, Phys.\ Rev.\ D \textbf{52}, 5729 (1995).

\bibitem{isz21} E.~Babichev and C.~Charmousis, JHEP \textbf{08}, 106 (2014).

\bibitem{isz22} S.~W.~Hawking, Nature \textbf{248}, 30 (1974).

\bibitem{isz23} S.~W.~Hawking, Commun.\ Math.\ Phys.\  \textbf{43}, 199 (1975) [erratum: Commun.\ Math.\ Phys.\  \textbf{46}, 206 (1976)].

\bibitem{isz24} J.~M.~Bardeen, B.~Carter and S.~W.~Hawking, Commun.\ Math.\ Phys.\  \textbf{31}, 161 (1973).

\bibitem{isz25} D.~N.~Page, Phys.\ Rev.\ Lett.\  \textbf{71}, 3743 (1993).

\bibitem{isz26} R.~M.~Wald, Living Rev.\ Rel.\  \textbf{4}, 6 (2001).

\bibitem{isz27} R.~A.~Konoplya and A.~Zhidenko, Rev.\ Mod.\ Phys.\  \textbf{83}, 793 (2011).

\bibitem{isz28} V.~Perlick, Living Rev.\ Rel.\  \textbf{7}, 9 (2004).

\bibitem{isz29} P.~V.~P.~Cunha and C.~A.~R.~Herdeiro, Gen.\ Rel.\ Grav.\  \textbf{50}, 42 (2018).

\bibitem{isz30} T.~Johannsen, Class.\ Quant.\ Grav.\  \textbf{33}, 124001 (2016).

\bibitem{isz31} A.~Abdujabbarov, M.~Amir, B.~Ahmedov and S.~G.~Ghosh, Phys.\ Rev.\ D \textbf{93}, 104004 (2016).

\bibitem{isz32} Z.~Younsi, A.~Zhidenko, L.~Rezzolla, R.~Konoplya and Y.~Mizuno, Phys.\ Rev.\ D \textbf{94}, 084025 (2016).

\bibitem{isz33} C.~Bambi and K.~Freese, Phys.\ Rev.\ D \textbf{79}, 043002 (2009).

\bibitem{isz34} K.~D.~Kokkotas and B.~G.~Schmidt, Living Rev.\ Rel.\  \textbf{2}, 2 (1999).

\bibitem{isz35} E.~Berti, V.~Cardoso and A.~O.~Starinets, Class.\ Quant.\ Grav.\  \textbf{26}, 163001 (2009).

\bibitem{isz36} H.~P.~Nollert, Class.\ Quant.\ Grav.\  \textbf{16}, R159 (1999).

\bibitem{isz37} V.~Ferrari and L.~Gualtieri, Gen.\ Rel.\ Grav.\  \textbf{40}, 945 (2008).

\bibitem{isz38} V.~Cardoso, E.~Franzin and P.~Pani, Phys.\ Rev.\ Lett.\  \textbf{116}, 171101 (2016) [erratum: Phys.\ Rev.\ Lett.\  \textbf{117}, no.8, 089902 (2016)].

\bibitem{isz39} T.~P.~Sotiriou and V.~Faraoni, Rev.\ Mod.\ Phys.\  \textbf{82}, 451 (2010).

\bibitem{isz40} S.~Capozziello and M.~De Laurentis, Phys.\ Rept.\  \textbf{509}, 167 (2011).

\bibitem{isz41} S.~A.~Hayward, Phys.\ Rev.\ Lett.\  \textbf{96}, 031103 (2006).

\bibitem{isz42} J.~M.~Bardeen, Proc.\ Int.\ Conf.\ GR5, Tbilisi, USSR (1968).

%\bibitem{isz43} Y.~S.~Myung and D.~C.~Zou, Eur.\ Phys.\ J.\ C \textbf{80}, 1036 (2020).

%\bibitem{isz44} J.~C.~S.~Neves, Phys.\ Rev.\ D \textbf{92}, 084015 (2015).

%\bibitem{isz45} A.~Övgün, Phys.\ Rev.\ D \textbf{105}, 084015 (2022).

\bibitem{isz46} N.~Dadhich, R.~Maartens, P.~Papadopoulos and V.~Rezania, Phys.\ Lett.\ B \textbf{487}, 1 (2000).

\bibitem{isz47} P.~S.~Letelier, Phys.\ Rev.\ D \textbf{20}, 1294 (1979).

%\bibitem{isz48} W.~A.~Hiscock, Phys.\ Rev.\ D \textbf{31}, 3288 (1985).

%\bibitem{isz49} B.~Linet, Gen.\ Rel.\ Grav.\  \textbf{17}, 1109 (1985).

\bibitem{isz50} I.~Dymnikova, Class.\ Quant.\ Grav.\  \textbf{19}, 725 (2002).

\bibitem{isz51} S.~N.~Solodukhin, Phys.\ Rev.\ D \textbf{51}, 609 (1995).

%\bibitem{isz52} M.~Barriola and A.~Vilenkin, Phys.\ Rev.\ Lett.\  \textbf{63}, 341 (1989).

%\bibitem{isz53} E.~F.~Eiroa, G.~E.~Romero and D.~F.~Torres, Phys.\ Rev.\ D \textbf{66}, 024010 (2002).

%\bibitem{isz54} A.~Borde, Phys.\ Rev.\ D \textbf{55}, 7615 (1997).

%\bibitem{isz55} L.~Balart and E.~C.~Vagenas, Phys.\ Rev.\ D \textbf{90}, 124045 (2014).


\bibitem{AHEP1} F. Ahmed, EPL {\bf 142}, 39002 (2023).

\bibitem{AHEP2} F. Ahmed, Int. J. Geom. Meths. Mod. Phys. {\bf 21}, 2450187 (2024).

\bibitem{AHEP3} B. Hamil, B C Lutfuoglu, F. Ahmed, Z. Yousaf, Nucl. Phys. B {\bf 1014}, 116861 (2025).



\bibitem{NPB} F.~Ahmed, A.~Al-Badawi, \.I.~Sakall\i{} and A.~Bouzenada, Nucl. Phys. B \textbf{1011}, 116806 (2025).

\bibitem{CJPHY} A. Al-Badawi and F. Ahmed, Chin. J. Phys. {\bf 94}, 185 (2025).

\bibitem{AHEP4} F. Ahmed, A. Al-Badawi, \.I. Sakall\i{}, and S. Kanzi, Phys. Dark Univ. {\bf 48}, 101907 (2025).

\bibitem{EPJC} F. Ahmed, J. Goswami and A. Bouzenada, Eur. Phys. J C {\bf 85}, 110 (2025).



\bibitem{VC} V. Cardoso, A. S. Miranda, E. Berti, H. Witek, and V. T. Zanchin, Phys. Rev. {\bf D 79}, 064016 (2009).

\bibitem{isz63} G.~W.~Gibbons and S.~W.~Hawking, Phys.\ Rev.\ D \textbf{15}, 2752 (1977).

\bibitem{isz64} V.~P.~Frolov, Phys.\ Rev.\ Lett.\  \textbf{115}, 051102 (2015).

%\bibitem{isz65} E.~T.~Akhmedov, T.~Pilling and D.~Singleton, Int.\ J.\ Mod.\ Phys.\ D \textbf{17}, 2453 (2008).

%\bibitem{frolov1} S.~Kala, H.~Nandan, K.~Maithani, S.~Roy and  A.~Abebe, arXiv:2503.19571 [gr-qc].

\bibitem{isz66} P.~C.~W.~Davies, Proc.\ Roy.\ Soc.\ Lond.\ A \textbf{353}, 499 (1977).

\bibitem{isz67} W.~G.~Unruh, Phys.\ Rev.\ D \textbf{14}, 870 (1976).

\bibitem{isz68} D.~V.~Fursaev, Phys.\ Rev.\ D \textbf{51}, 5352 (1995).

\bibitem{isz69} S.~B.~Giddings, Phys.\ Rev.\ D \textbf{74}, 106005 (2006).

\bibitem{isz70} J.~D.~Bekenstein, Phys.\ Rev.\ D \textbf{7}, 2333 (1973).

\bibitem{isz71} S.~Carlip, Int.\ J.\ Mod.\ Phys.\ D \textbf{23}, 1430023 (2014).

\bibitem{isz72} S.~W.~Hawking, arXiv:1401.5761 [hep-th].

\bibitem{isz73} P.~C.~W.~Davies, Class.\ Quant.\ Grav.\  \textbf{6}, 1909 (1989).

\bibitem{isz74} R.~M.~Wald, Living Rev.\ Rel.\  \textbf{4}, 6 (2001).

\bibitem{isz75} S.~W.~Hawking, Phys.\ Rev.\ D \textbf{14}, 2460 (1976).

\bibitem{isz76} A.~Almheiri, D.~Marolf, J.~Polchinski and J.~Sully, JHEP \textbf{02}, 062 (2013).

\bibitem{isz77} S.~W.~Hawking, M.~J.~Perry and A.~Strominger, Phys.\ Rev.\ Lett.\  \textbf{116}, 231301 (2016).

\bibitem{isz81} A.~Abdujabbarov, M.~Amir, B.~Ahmedov and S.~G.~Ghosh, Phys.\ Rev.\ D \textbf{93}, 104004 (2016).

\bibitem{isz82} Z.~Younsi, A.~Zhidenko, L.~Rezzolla, R.~Konoplya and Y.~Mizuno, Phys.\ Rev.\ D \textbf{94}, 084025 (2016).

\bibitem{isz83} J.~L.~Synge, Mon.\ Not.\ Roy.\ Astron.\ Soc.\  \textbf{131}, 463 (1966).

\bibitem{isz84} S.~Chandrasekhar, The Mathematical Theory of Black Holes, (Oxford University Press, Oxford (UK), 1983).

\bibitem{isz86} J.~Chagoya, C.~Ortiz, B.~Rodr\'\i{}guez and A.~A.~Roque, Class. Quant. Grav. \textbf{38}, 075026 (2021).

\bibitem{isz87} P.~V.~P.~Cunha, C.~A.~R.~Herdeiro and E.~Radu, Phys.\ Rev.\ D \textbf{96}, 024039 (2017).

\bibitem{isz88} H.~Falcke, F.~Melia and E.~Agol, Astrophys.\ J.\ Lett.\  \textbf{528}, L13 (2000).

\bibitem{isz89} T.~Johannsen, Astrophys.\ J.\  \textbf{777}, 170 (2013).

\bibitem{isz90} A.~de Vries, Class.\ Quant.\ Grav.\  \textbf{17}, 123 (2000).

\bibitem{isz91} D.~Psaltis et al. [Event Horizon Telescope Collaboration], Phys.\ Rev.\ Lett.\  \textbf{125}, 141104 (2020).

\bibitem{isz92} P.~V.~P.~Cunha and C.~A.~R.~Herdeiro, Gen.\ Rel.\ Grav.\  \textbf{50}, 42 (2018).

%\bibitem{isz93} R.~Takahashi, Astrophys.\ J.\  \textbf{611}, 996 (2004).

%\bibitem{isz94} A.~Bonanno and S.~Silverava, Phys. Rev. D \textbf{99}, 101501 (2019).

\bibitem{isz95} A.~F.~Zakharov, Phys.\ Rev.\ D \textbf{90}, 062007 (2014).

\bibitem{isz96} T.~Johannsen and D.~Psaltis, Astrophys.\ J.\  \textbf{718}, 446 (2010).

%\bibitem{isz97} S.~Doeleman, Nature Astron. \textbf{1}, 646 (2017).

\bibitem{isz98} E.~Barausse, V.~Cardoso and P.~Pani, Phys. Rev. D \textbf{89}, 104059 (2014).

\bibitem{d3} S. Iyer, C.M. Will, Phys. Rev. D \textbf{35}, 3621 (1987).

\bibitem{d4} R.A. Konoplya, Phys. Rev. D \textbf{68}, 024018 (2003).

\bibitem{isz99} F.~Rahaman, R.~Amin, M.~Hasan, A.~Islam, S.~Ray, A.~Aziz and N.~A.~Pundeer, Fortsch. Phys. \textbf{73}, 2400007 (2025).

\bibitem{isz98x} S.~W.~Hawking, arXiv:1401.5761 [hep-th].

\bibitem{isz98xx} B.~Pourhassan, X.~Shi, S.~S.~Wani, Saif-Al-Khawari, F.~Kazemian, \.I. Sakall\i{}, N.~A.~Shah and M.~Faizal, JHEP \textbf{25}, 109 (2025).

\bibitem{isz99x} P.~C.~W.~Davies, Class.\ Quant.\ Grav.\  \textbf{6}, 1909 (1989).

\bibitem{isz100x} E.~Berti, V.~Cardoso and A.~O.~Starinets, Class.\ Quant.\ Grav.\  \textbf{26}, 163001 (2009).

\bibitem{isz101x} R.~P.~Kerr, Phys.\ Rev.\ Lett.\  \textbf{11}, 237 (1963).

\bibitem{isz102x} B.~P.~Abbott et al. [LIGO Scientific and Virgo Collaborations], Phys.\ Rev.\ Lett.\  \textbf{116}, 061102 (2016).


\end{thebibliography}
\end{document}